\documentclass[aps, prb, amsmath, amssymb, reprint, floatfix, superscriptaddress, longbibliography]{revtex4-2}

\usepackage[utf8]{inputenc}
\usepackage[T1]{fontenc}

\usepackage{amsfonts}
\usepackage{stmaryrd}
\usepackage{graphicx}
\usepackage{siunitx}
\usepackage{xcolor}
\usepackage{physics} 
\usepackage[colorlinks=true, citecolor=blue, urlcolor=blue]{hyperref}
\usepackage{orcidlink}

\newcommand{\Ttilde}{\widetilde{T}}
\newcommand{\Tbar}{\overline{T}}
\newcommand{\aPtilde}{\widetilde{aP}}
\newcommand{\Ptilde}{\widetilde{P}}
\newcommand{\atilde}{\widetilde{a}}
\newcommand{\aPbar}{\overline{aP}}
\newcommand{\abar}{\overline{a}}
\newcommand{\Pbar}{\overline{P}}
\newcommand{\thetatilde}{\widetilde{\theta}}
\newcommand{\ba}{\mathrm{ba}}
\newcommand{\e}{\mathrm{e}}

\begin{document}

\title{Temperature mediated back-action in micro- and nanomechanical resonators}

\author{L. Bellon\,\orcidlink{0000-0002-2499-8106}}
\email{ludovic.bellon@ens-lyon.fr}
\affiliation{\href{https://ror.org/02feahw73}{CNRS}, \href{https://ror.org/04zmssz18}{ENS de Lyon}, \href{https://ror.org/00w5ay796}{Laboratoire de Physique}, F-69342 Lyon, France}

\author{P. Verlot\,\orcidlink{0000-0002-5105-3319}}
\affiliation{Université Paris-Saclay, CNRS, ENS Paris-Saclay, CentraleSupélec, LuMIn, 91405, Orsay, 
France}
\affiliation{Institut Universitaire de France, 1 rue Descartes, 75231 Paris, France}

\date{\today}

\begin{abstract}
We theoretically investigate the thermally induced back-action effects in absorption-sensitive micro- and nanomechanical resonators. We propose a unified approach, enabling to simultaneously address both the effective dynamics and non-equilibrium phononic state, depending on the position of a punctual sensing (and heating) probe at the surface of the mechanical device. We present an analytical solution in terms of green functions for a unidimensional resonator whose thermomechanical deformation profile generally follows that of the mechanical losses. In particular, we find that both the dynamics and the mechanical fluctuations strongly depend on the loss distribution. The effect of the quantum fluctuations of the heat source is also discussed. Our approach provides the first steps towards a thorough, general platform for analyzing thermal back-action effects and their consequences, which may be of significance for future development in ultrasensitive nanomechanical research.
\end{abstract}

\maketitle 

\section{Introduction}

Micro- and nanomechanical resonators have recently surged as a prominent class of ultrasensitive probes, both in Science and Technology~\cite{o2010quantum, Gil-Santos2010, chan2011laser, Liu-2021}: Owing to the unique combination of a ultra-low mass and high mechanical Q-factor, these devices have the ability to convert very small external driving phenomena into large displacements, which are to be further readout with a high precision, using a localized high energy probe~\cite{aspelmeyer2014cavity}. 

An essential task in mechanical sensing is to accurately determine both the mechanical response and mechanical noise, the former bridging the measured motion to the driving phenomenon of interest, while the latter setting the sensitivity of the measurement~\cite{Saulson1990}. Due to their reduced size and high aspect ratio, micro- and nanomechanical transducers generally display higher thermal resistances, resulting in much enhanced responses to thermal variations~\cite{blaikie2019fast, Chen-2022}. In this context, the impact of absorption-induced thermal gradients has long been known through the mechanism of dynamical back-action~\cite{braginski1967ponderomotive}. The principle is for thermal gradients to create a strain-dependent stress inside the resonator, which feeds back to the deformation, resulting in both a modified effective mechanical response and phonon occupancy~\cite{metzger2004cavity, verhagen2012quantum,Nigues2014a}. While providing a successful, qualitative description of the dynamics associated with thermal gradients, this approach often relies on coarse-grained descriptions of the thermal-to-motion conversion mechanisms~\cite{Metzger2008, tavernarakis2018optomechanics, fogliano2021mapping}. More specific models have been introduced~\cite{ramos2012optomechanics, primo2021accurate}, but are based on numerical simulation describing particular resonators, and have not been extended to generic cases. Most importantly, the existing approaches leave aside a fundamental aspect related to the presence of thermal gradient, that is the out-of-equilibrium nature of mechanical states subjected to thermal gradients~\cite{Geitner2017, shaniv2023direct, Collin2023}. 

In this paper, we propose a unified theoretical approach that describes both the dynamical response and the fluctuation distribution of a nanomechanical system crossed by a large, measurement-induced thermal gradient. We specifically present an analytical solution of a unidimensional resonator, whose thermomechanical deformation profile is generally assumed to follow its thermoelastic losses. The resulting thermal back-action force is derived and computed as a function of the position of a punctual probe along the resonator, and the associated fluctuations are analyzed, both in the thermal and quantum regimes. Our work aims at representing the first steps towards a general, thermal back-action analysis platform for ongoing and future ultrasensitive nanomechanical research.

\section{General hypothesis}
We consider a one-dimensional mechanical resonator, such as a cantilever, beam or rod, operated in vacuum. Typical examples include micro-cantilevers~\cite{Lavrik2004} (such as those used for atomic force microscopy~\cite{Garcia2002}) and micro- and nanowires~\cite{rossi2017vectorial, de2017universal, tavernarakis2018optomechanics} (NW). We will use the latter denomination in this article without loss of generality. The NW can oscillate in the transverse direction $x$. $\xi(y)$ depicts its small amplitude deflection as a function of its longitudinal coordinate $y$. The computation is performed in the clamped-free geometry, but extension to double-clamped NW is straightforward. We limit ourselves to the detection scheme where a tightly confined probe beam is used to measure the deflection in $y=y_p$, e.g. a laser~\cite{ramos2012optomechanics, rossi2017vectorial, de2017universal, tavernarakis2018optomechanics} or an electron beam~\cite{Buks2000, Nigues2014a, pairis2019shot, cretu2022direct}. This probe is focused in $(x_p, y_p)$, in a fashion such that the transverse vibrations produce a modification of the beam-NW interaction that can be detected by a suitable sensor (e.g. split photodetector~\cite{sanii2010high}, interferometric sensor~\cite{paolino2013quadrature}, secondary electron detector~\cite{pairis2019shot}, etc.). As a side effect of the detection scheme, a fraction $a$ of the incoming probe beam power $P$ is absorbed, resulting in a temperature rise of the NW. We are interested here in computing the back-action forces associated with this process and how it affects the resonance~\cite{metzger2004cavity, Metzger2008, DeLiberato2011, ramos2012optomechanics, tsioutsios2017real}. Throughout this article, we always assume small amplitude fluctuations so as to work in a linear framework.

\begin{figure}[tb]
\begin{center}
\includegraphics[width=85mm]{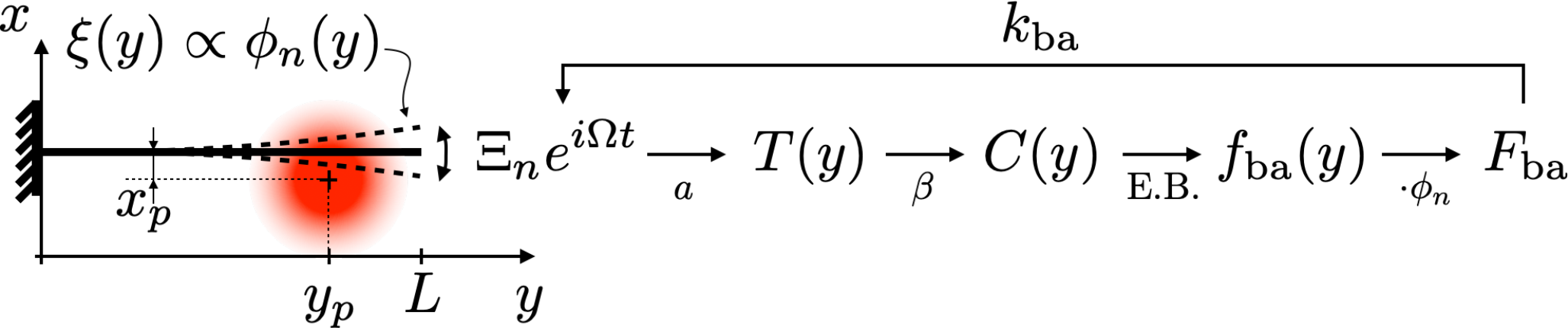}
\caption{Methodology: we consider a 1D mechanical oscillator of length $L$ along the $y$ axis, vibrating along the transverse $x$ axis at its resonance angular frequency $\omega$, with an amplitude $\Xi_n\phi_n(y)$ proportional to the eigenmode shape $\phi_n$. This motion is probed with a light or electron beam focused close to its edge, sketched here with a gaussian shaped red spot centered in $(x_p, y_p)$. A fraction $a$ of the beam energy is absorbed by the resonator, leading to a dynamic temperature field $T(y)$. This field changes the equilibrium curvature $C(y)\sim\beta T(y)$, which in turns creates a stress distribution. Using the Euler-Bernoulli description of the resonator, this stress is equivalent to an external distributed back-action force $f_\ba(y)$, which we project on $\phi_n(y)$ to compute the lumped force $F_\ba$ acting on the resonant mode. Since this force is proportional to the original mode amplitude $\Xi_n$, we can define the associated back-action stiffness $k_\ba$, that embeds how the resonance is affected by the probe beam.}
\label{Figconcept}
\end{center}
\end{figure}

We use the methodology described in Fig.~\ref{Figconcept}. We start by imposing an oscillation amplitude $\Xi_n$ at the angular frequency $\omega$, close to the resonance $\omega_n$ corresponding to the eigenmode $\phi_n(y)$. The corresponding deflection is given by $\xi(y, t)=\Xi_n \phi_n(y)e^{i\omega t}$. The distance $\xi(y_p, t)-x_p$ between the probe beam and the oscillator is therefore modulated, as is the absorbed power $aP$. This motion thus creates a dynamic temperature field $T(y)$, that in turns creates a distributed force $f_\ba(y)$ along the NW, which we project on the eigenmode shape $\phi_n(y)$ to infer the force $F_\ba$ acting on the resonant mode. This back-action force is proportional to the imposed deflection amplitude $\Xi_n$, with the proportionality constant $k_\ba$ embedding its effect on the resonance. The signature of this interaction is the modification of the resonance frequency (real part of the stiffness $k_\ba$), and of the damping (imaginary part of $k_\ba$, in quadrature with the oscillation). The methodology is conceptually similar to Refs.~\onlinecite{ramos2012optomechanics, primo2021accurate}, which are each dedicated to a very specific sample and detection. Here we develop a very broad framework that explores generic resonators and different classes of optomechanical coupling, and will demonstrate the dramatic effect of the probe position.
The following sections propose an analytical framework to implement this methodology. In Section \ref{section:heat}, we compute the dynamic temperature field due to the absorption of the probe beam. In Section \ref{section:Fba}, we translate this temperature distribution into the back-action force acting on the oscillation mode. In section \ref{section:kba}, we present several hypotheses for the stress generated by the temperature field, and their implication for the back-action stiffness $k_\ba$, and thus for the resonance. In the next two sections, we explore the effect of the back-action force noise associated with the quantum fluctuations of the probe beam (section \ref{section:shotnoise}), and the additional, classical mechanical thermal noise component acting on the system (\ref{section:thermalnoise}). We then compare the relative contribution of these two sources of noise in a precision force measurement scheme in section \ref{section:forcenoise}. Finally, we present a numerical application to a typical mechanical nanowire, highlighting the orders of magnitude of the various effects in \ref{section:appcase}, before the conclusions and perspectives of this work.

\section{Heat equation} \label{section:heat}

We assume that the transverse directions of the resonator are much smaller than any relevant length scale, so that a one-dimensional description of the fields is sufficient. The nanowire radius $R$ is therefore much smaller that its length $L$. Since vacuum is assumed, any heat exchange with the surroundings (including thermal radiation) is neglected. The heat equation for the temperature field $T(y, t)$ along the nanowire is:
\begin{equation}
\rho c \partial_t T - \partial_y (\kappa \partial_y T) = \frac{a p}{\pi R^2}
\end{equation}
with $\kappa$ the thermal conductivity, $\rho$ the density, $c$ the heat capacity of the oscillator, and $ap(y,t)$ the power influx density due to the probe beam partial absorption. Unless otherwise noted, we work under the hypothesis that the probe beam spot is small and can be described as punctual, thus this forcing term is modeled by a Dirac distribution $ap(y,t)=aP(t)\delta^D(y-y_p)$. The boundary conditions (BC) correspond to a fixed temperature $T_0$ at the origin (that is the clamping position $y=0$) and no heat flux at the nanowire free end ($y=L$):
\begin{subequations}
\begin{align}
T(0, t)&=T_0, \\
\kappa\partial_y T(L, t)&=0 \label{eqBCthL}
\end{align}
\end{subequations}
Assuming $\kappa$ is uniform, the static temperature profile $\Tbar$ solution of these equations is piecewise linear:
\begin{equation} \label{eq.Tbar}
\Tbar(y)=T_0+\bar \theta \frac{1}{L}\min(y, y_p), 
\end{equation}
where $\bar \theta = \aPbar L/(\pi R^2 \kappa)$ is the absolute maximum static temperature elevation (corresponding to $y=y_p=L$), and $\aPbar$ is the mean absorbed power from the probe beam. The maximum power one can use for a given oscillator is constrained by this maximum temperature, as too large values may damage the system.

We now define $\Ttilde=T-\Tbar$ as the temperature deviation from this stationary state, and study the dynamics of the temperature field due to the non-stationary part of the absorbed power $\aPtilde=aP-\aPbar$. In the Fourier space, the heat equation writes:
\begin{equation} \label{EqDynHeat}
\alpha^2 \Ttilde - \partial_y^2 \Ttilde = \frac{\aPtilde(\omega)}{\pi R^2 \kappa}\delta^D(y-y_p), 
\end{equation}
with
\begin{equation} \label{alphaDynHeat}
\alpha=\sqrt{i\omega \frac{\rho c}{\kappa}} = \e^{i\pi/4}\sqrt{\omega \frac{\rho c}{\kappa}}
\end{equation}
and the dynamic BCs $\Ttilde(0, \omega)=0$ and $\partial_y \Ttilde(L, \omega)=0$. We solve this equation with the ansatz $\Ttilde = A \sinh\alpha y$ for $y<y_p$ and $\Ttilde = B \cosh\alpha (y-L)$ for $y>y_p$ (satisfying the heat equation and the BCs). The continuity of the temperature field in $y=y_p$ imposes:
\begin{equation}
A \sinh \alpha y_p = B \cosh \alpha (y_p-L), 
\end{equation}
while the discontinuity of the heat flux due to the Dirac source term writes:
\begin{equation}
A \alpha \cosh \alpha y_p - B \alpha \sinh \alpha (y_p-L)= \frac{\aPtilde}{\pi R^2 \kappa}.
\end{equation}
We solve these two equations for $A$ and $B$ and conclude that the dynamic temperature field is given by:
\begin{align}
\Ttilde(y, \omega) &= \thetatilde(\omega) h\left(y, \sqrt{i\omega\rho c/\kappa}\right), \\[1mm]
\thetatilde(\omega) &= \frac{L}{\pi R^2 \kappa}\aPtilde(\omega) , \\[1mm]
\begin{split}
h(y, \alpha)&= H(y_p-y) \frac{\sinh \alpha y  \cosh \alpha (y_p-L)}{\alpha L \cosh \alpha L}\\
& \quad +H(y-y_p)\frac{\sinh \alpha y_p  \cosh \alpha (y-L)}{\alpha L \cosh \alpha L}.
\end{split}
\end{align}
The dynamic temperature field is expressed as the product of 2 terms, $\thetatilde(\omega)$ with quantifies the amplitude of the forced temperature variation, and $h(y, \alpha)$ (where $\alpha$ is the complex number proportional to $\sqrt{i\omega}$ defined in Eq.~\ref{alphaDynHeat}) which accounts for the propagation of the damped thermal wave according to the heat equation. Note that all three quantities expressed in Eqs. 8 to 10 are implicit variables of the position $(x_p, y_p)$ of the probe beam. 

\begin{figure}[htbp]
\begin{center}
\includegraphics{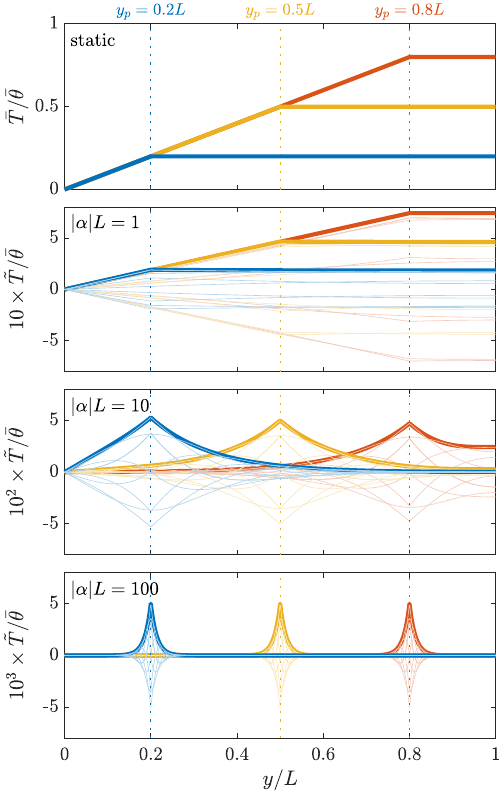}
\caption{Temperature field along the 1D resonator for a point like source located in $y_p$. The top panel reports the static profile $\bar T$ assuming a constant thermal conductivity, normalized to the maximum temperature one would get by choosing $y_p=L$, for 3 values of $y_p$ reported above the plots. The  bottom panels present the dynamic profile $\tilde T$ for three values of thermal diffusion length $1/|\alpha|$. $|\alpha|L=1$ is the low frequency / high conductivity limit. The thick line corresponds to the envelope of the thermal waves (thin lines at different times) inside the NW, and the colors match the static profiles corresponding to the 3 values of $y_p$. As expected, this low frequency limit is equivalent to the static field. $|\alpha|L=10$ is the intermediate thermal diffusivity regime. The thermal waves travel with a complex shape across the nanowire, with an asymmetric profile reflecting that of the clamped-free boundary conditions. $|\alpha|L=100$ is the high frequency~/~low conductivity limit. The thermal waves are localized around the heating point $y_p$. The amplitude of all dynamic fields is computed with a $\SI{100}{\%}$ modulation depth of the static profile ($\aPtilde=\aPbar$). The lower the thermal length, the lower the temperature wave amplitude (note the multiplicative factors on the vertical scales). This effect is usually noticed as a high frequency cutoff of thermal actuation of micro-oscillators~\cite{Allegrini-1992, metzger2004cavity}. }
\label{Tprofile}
\end{center}
\end{figure}

Those temperature fields, static and dynamic, are reported in Fig.~\ref{Tprofile}. The value of $|\alpha| L$ sets the regime to be considered: the dynamic profile is equivalent to the static one in the low frequency / high conductivity limit ($|\alpha| L\ll 1$), and tends to be very localized and of low amplitude in the opposite limit ($|\alpha| L\gg 1$).

As those fields are computed with a Dirac source term, they represent the Green functions associated to the linear heat equation. They can be used to compute the temperature profile created by a distributed probe beam heating. In such a case, the curves in Fig.~\ref{Tprofile} should be convoluted to the power influx density $ap(y,\omega)$. As long as the spatial extension of the probe beam is small compared to $1/|\alpha|$, the temperature profiles of Fig.~\ref{Tprofile} are representative, and the punctual hypothesis holds. In the following, we work under this specific hypothesis and thus derive the green functions for the feedback mechanism, leaving the convolution step as a perspective shortly discussed in section \ref{sec:ccl}.

\section{Temperature mediated force back-action mechanism}  \label{section:Fba}

As the temperature of the nanowire varies, it can induce stresses that result in a change in the local curvature of equilibrium $C=\partial_y^2 \xi$. For example, let us assume the local curvature at temperature $T_0$ is $C_0\neq 0$. When the temperature changes to $T$, the equilibrium curvature changes to $C=C_0/[1+\varepsilon(T-T_0)]\sim C_0[1-\varepsilon(T-T_0)]$, with $\varepsilon$ the linear thermal expansion coefficient. In other cases, for instance when several materials are involved in the oscillator composition, a bimetal effect can be present~\cite{ramos2012optomechanics}, which also results in an equilibrium curvature depending on the local temperature. Yet in other cases, some defects or geometric asymmetries may be present and trigger similar effects. To describe the generic case, we assume that these curvature effects are small and define a linear thermal bending coefficient $\beta$, so that in equilibrium:
\begin{equation}
C = C_0+ \beta (T-T_0). 
\end{equation}
The equation of motion of the nanowire is a variant of Euler-Bernoulli's, and writes in Fourier space:
\begin{align}
\!\!\!\!\partial_y^2 [EI(\partial_y^2 \xi - \beta \Ttilde)] - \rho \pi R^2 \omega^2 \xi & = f_\mathrm{ext}, \\
\partial_y^2 [EI\partial_y^2 \xi] - \rho \pi R^2  \omega^2 \xi & = \partial_y^2 (EI \beta \Ttilde)+f_\mathrm{ext}, \label{EqEBcoupledT}
\end{align}
where static terms have been removed~\footnote{Just as the static temperature field depends linearly on the mean power $\Pbar$, the static deflection profile $\bar \xi(y)$ can change due to the static temperature field $\Tbar(y)$. This effect will just offset the average position of the oscillator, and is taken into account by defining $\xi$ as the dynamic deviation to $\bar \xi$. When tuning the lateral position $x_p$ of the probe to maximize sensitivity, one should recover the same level of absorption $a$ for any imposed $\Pbar$. In the linear framework that we use, this static deflection does not have any effect on the dynamics.}, $EI$ is the flexural rigidity of the nanowire, and $f_\mathrm{ext}$ a potential external distributed force. The temperature field can thus be seen as such an external force field, 
\begin{align}
f_\ba (y, \omega) &= \partial_y^2 [EI \beta(y) \Ttilde(y, \omega)], \\
& = EI \thetatilde(\omega) \partial_y^2 [\beta(y) h(y, \sqrt{i\omega\rho c/\kappa})].
\end{align}

We suppose that its effect on the mechanical response of the nanowire is small, so that the resonant mode shape $\phi_n(y)$ is unaffected by $f_\ba$. We can then project Eq.~\ref{EqEBcoupledT} on the solution of the unperturbed Euler-Bernoulli equation $\phi_n(y)$, and derive in the Fourier space the equation of the driven simple harmonic oscillator corresponding to the resonant mode:
\begin{equation} \label{eq.OH}
(k^r_n+ik^i_n - m \omega^2) \Xi_n(\omega) = F_\ba(x_p, y_p, \omega) + F_\mathrm{ext}.
\end{equation}
Here $m=\rho R^2L$ is the mass of the oscillator, $k^r_n$ is the lumped stiffness of mode $n$, and $k^i_n$ describes the damping, accounted for in Eq.~\ref{EqEBcoupledT} by the imaginary part $E^i$ of Young's Modulus $E=E^r+iE^i$:
\begin{equation}
k^r_n+ik^i_n = \int_{0}^{L} (E^r+iE^i) I \phi_n''(y)^2 \dd y.
\end{equation}
Finally, the temperature mediated force back-action is:
\begin{align}
F_\ba(x_p, y_p, \omega) &=\int_0^L f_\ba(y, \omega) \phi_n(y) \dd y, \\
&=  \aPtilde(x_p, y_p, \omega) \frac{L}{\pi R^2 \kappa} g(y_p, \omega), \label{EqFba}
\end{align}
with
\begin{equation}
\!\!\!g(y_p, \omega) = \int_0^L\partial_y^2 [EI \beta(y) h(y, \sqrt{i\frac{\omega \rho c}{\kappa}}, y_p)] \phi_n(y) \dd y. \label{Eqg}
\end{equation}
In those last equations, we explicitly expressed the dependence of the back-action force with respect to the position $(x_p, y_p)$ of the probe beam.

There are two origins for the variations of $\aPtilde$ that drive the dynamic temperature field. On the one hand, the probe beam power intrinsically fluctuates in time, yielding to first order to $\Ptilde(\omega)\neq 0$ and $\aPtilde=\abar\Ptilde$ (with $\abar$ the static absorption). On the other hand, the nanowire is laterally deflected with respect to the probe beam, changing the dynamic absorption $\atilde$:
\begin{subequations}
\begin{align} 
\atilde(x_p, y_p, \omega)&=\frac{\dd a}{\dd x_p}(x_p)  \xi(y_p, \omega)\\
&= \frac{\dd a}{\dd x_p}(x_p) \phi_n(y_p) \Xi_n(\omega). \label{Eqatilde}
\end{align}
\end{subequations}
In the following sections, we independently treat each of these two mechanisms.

\section{Back-action effects on the oscillator mechanical response}  \label{section:kba}

\begin{figure*}[tb]
\begin{center}
\includegraphics{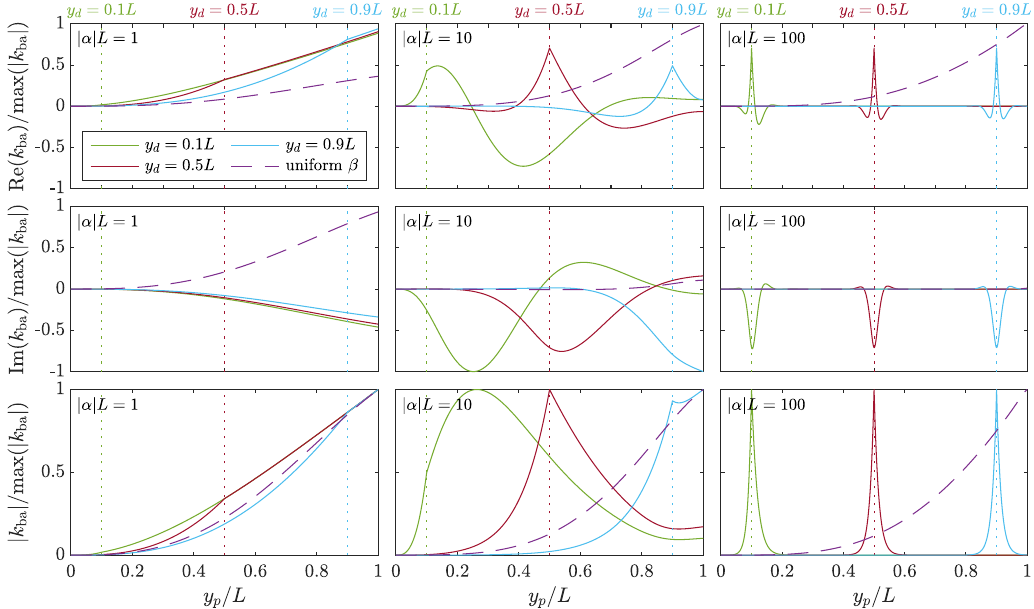}
\caption{Back-action stiffness spatial profile for different spatial distributions of $\beta$. The top row presents the real part $k_\ba^r$ of the stiffness (modifying the resonance frequency), the middle one the imaginary part $k_\ba^i$ (modifying the damping), and the bottom one the modulus $|k_\ba|$. The three columns correspond to the thermal diffusivity length $|\alpha| L$ reported in the upper left corner of the plots. The curves are normalized to the maximum of $|k_\ba(y_p)|$ on all positions $y_p$. The dashed curve corresponds to a uniform $\beta$ (Eq.~\ref{EqgBetaUniform}), the plain ones to a single defect in $y_d/L=0.1, 0.5$ or $0.9$ (Eq.~\ref{EqgBetaPeaked}, assuming $\tau_d=0$). All curves are computed for the first oscillation mode of the nanowire, when the position $y_p$ of the probe beam is scanned along the resonator length. The signature of the temperature mediated back-action is very different for the various hypotheses probed here, and displays a non-trivial shape, including sign changes in some cases.}
\label{Figk_ba}
\end{center}
\end{figure*}

In this section, we focus on the driving effect resulting from the temperature field induced by the vibration of the nanowire: $\aPtilde=\atilde \Pbar \propto \Xi_n(\omega)$. This vibration creates a thermal wave, which in turn drives the motion of the nanowire via the above-described thermomechanical coupling. The fields of deformation and of temperature are thus coupled, and the induced feedback modifies the effective mechanical response of the system due to the presence of the measurement probe beam. This coupling can cause self-oscillation or back-action cooling of the oscillator~\cite{Metzger2008, metzger2004cavity, ramos2012optomechanics, barton2012, Nigues2014a, tsioutsios2017real}.

Injecting the expression of $\atilde$ from Eq.~\ref{Eqatilde} into Eq.~\ref{EqFba}, we can define a back-action stiffness $k_\ba$ with
\begin{align}
F_\ba(x_p, y_p, \omega) &= - k_\ba (x_p, y_p, \omega) \Xi_n(\omega), \\
k_\ba (x_p, y_p, \omega) & = - \frac{\dd a}{\dd x_p}(x_p)  \Pbar \frac{L}{\pi R^2 \kappa} \phi_n(y_p) g(y_p, \omega). \label{Eqkba}
\end{align}
This back-action mechanical stiffness is a complex quantity $k_\ba=k_\ba^r+ik_\ba^i$. To describe the dynamics of the equivalent resonator, Eq.~\ref{eq.OH} therefore has to be modified in:
\begin{equation} \label{eq.OHba}
[k^r_n+k_\ba^r +i(k^i_n+k_\ba^i)- m \omega^2] \Xi_n (\omega)= F_\mathrm{ext}.
\end{equation}
The real part $k_\ba^r$ changes the effective stiffness of the oscillator, thus the resonance frequency, whereas the imaginary part $k_\ba^i$ modifies the damping.

A first feature of $k_\ba$ in this framework is that it is linear in the probe power $\Pbar$.  A noticeable second feature is that $k_\ba$ is proportional to $\dd a / \dd x_p$, which is antisymmetric with respect to the lateral position $x_p$. The frequency shift and the damping modification resulting from the back-action should be opposite on the left and right sides of the nanowire. This essentially stems from the assumption that the heat-induced mechanical deformation is purely dependent on the absorbed power. In particular, by choosing two symmetric points $\pm x_p$, the average of the two complex stiffnesses cancels the back-action effect, such that both the intrinsic stiffness and the damping of the resonator can be recovered. The two effects (power dependency and antisymmetry) have been experimentally reported, where the back-action typically results in amplification or cooling of mechanical modes depending on the side sensed by the probe, proportionally to the power implied~\cite{Nigues2014a, tavernarakis2018optomechanics}.

To go further, additional hypotheses are needed, regarding the spatial distribution of the thermo-elastic deformation $EI\beta$. For example, for a nanowire with a uniform natural curvature $C_0$, the linear thermal bending coefficient is uniform: $\beta = -C_0 \varepsilon$. The same uniformity would also apply to a nanowire coated on one side only~\cite{ramos2012optomechanics} and thus subject to the ``bimetal effect'' (that is, the differential expansion of its layers). In such cases, we extract $EI\beta$ from the integral in Eq.~\ref{Eqg} and get:
\begin{equation}
g(y_p, \omega) = i EI \beta \omega \frac{ \rho c}{\kappa}\int_0^L  h(y, \sqrt{i \frac{\omega \rho c}{\kappa}}, y_p) \phi_n(y)\dd y. \label{EqgBetaUniform}
\end{equation}
The dependency on the longitudinal coordinate $y_p$ of the back-action force can be estimated from this formula since the functions $h$ and $\phi_n$ are known.

Another relevant hypothesis on $\beta$ is that it corresponds to a localized, point-like defect: $\beta(y, \omega)=\beta_d e^{i\omega \tau_d} \delta^D (y-y_d)$, with $\delta^D$ the Dirac distribution centered at the position $y_d$ of the defect, $\beta_d$ the amplitude of the punctual thermal bending coefficient, and $\tau_d$ a microscopic response time of the mechanical stress to the temperature change. After a double integration by parts, Eq.~\ref{Eqg} leads to:
\begin{equation}
g(y_p, \omega) = EI\beta_d e^{i\omega \tau_d} h(y_d, \sqrt{i\frac{\omega \rho c}{\kappa}}, y_p) \phi_n''(y_d). \label{EqgBetaPeaked}
\end{equation}
Here again, the spatial dependency of the back-action effect on $y_p$ can be predicted and compared to the experimental data. If several defects are present, all that is required is to sum the functions $g$.

The effect of the spatial distribution of $\beta$ on the spatial evolution of the back-action stiffness $k_\ba$ is illustrated in Fig.~\ref{Figk_ba}, using for the mechanical mode shape $\phi_1(y)$, the first normal mode of a clamped-free nanowire. Other modes could be computed using the very same recipe. The different hypotheses yield very different signatures both for the real and imaginary part. Experimental observations on the resonance frequency shift and the damping variation, both along and across the nanowire, will thus point towards the best model to describe a specific thermal bending mechanism. These quantities can be extracted either from measuring the response function to an external drive (\emph{i.e.} by applying an external force $F_\mathrm{ext}$ and measuring the deflection while sweeping the driving frequency), or directly from a thermal noise measurement. Indeed, providing a high enough motion readout sensitivity, the spontaneously observed Brownian fluctuations directly lead to the resonance characteristics, with a predictable, antisymmetric behavior from one lateral position of the probe beam to the opposite: By simply changing the sign of $x_p$, the effect of the thermal back-action can be switched from an upward to a downward frequency shift and from a cooling to an amplification of the noise amplitude. In particular, the case described in Fig.~\ref{Figk_ba} corresponds to a reduced damping, stemming from the arbitrary convention $(\dd a/\dd x_p)_{x_p}>0$ which yields a negative imaginary part $k_\ba^i$, associated with motion amplification, and which can even lead to self-oscillations if the negative back-action damping overcomes the intrinsic dissipation~\cite{barton2012, Nigues2014a, tsioutsios2017real}. This effect can be reversed, by symmetrically moving the measurement probe, resulting in motion cooling~\cite{Nigues2014a, tavernarakis2018optomechanics}. Note that such antisymmetries essentially reflect those displayed by the mechanical structure in response to heat, by nature, or by design. Their presence therefore enables to unequivocally conclude as to the nature of the observed measurement back-action force, specifically when the expected fundamental mechanism has distinct symmetries (e.g., scattering forces~\cite{ashkin1986observation, jain2016direct}, gradient forces~\cite{anetsberger2009near}...). Likewise, our approach can be straightforwardly extended to other, defect-sensitive measurement processes~\cite{arcizet2011single, teissier2014strain, kettler2021inducing}.

\section{Quantum back-action from power fluctuations}  \label{section:shotnoise}

In this section, we focus on the variations of $\aPtilde$ due to the power fluctuations: $\aPtilde = \abar \Ptilde$. Even for a stabilized input beam, such fluctuations are unavoidably present due to the quantum nature of the measurement probe e.g., photon-shot noise, electron-shot noise, etc. These quantum fluctuations exhibit a white power spectrum:
\begin{equation}
S_P= \frac{\langle \Ptilde \rangle}{\Delta \omega} = 2\epsilon \Pbar, 
\end{equation}
with $\Delta \omega$ the spectral bandwidth, $\epsilon$ the quantum of energy of the particle involved ($\epsilon=h\nu$ for a photon of frequency $\nu$, with $h$ the Plank constant, $\epsilon=eV$ for an electron of charge $-e$ and acceleration voltage $V$).

Eq.~\ref{EqFba} simply leads to:
\begin{align}
F_\ba^\mathrm{SN}(x_p, y_p, \omega) &=  \abar (x_p) \Ptilde(\omega) \frac{L}{\pi R^2 \kappa} g(y_p, \omega), \label{EqFbaQ}
\end{align}
where the superscript $^\mathrm{SN}$ stand for ``shot noise'' driven. The measured deflection $\xi(y_p, \omega)$ is deduced from the transfer function of the harmonic oscillator, yielding
\begin{equation}
\xi^\mathrm{SN}(y_p, \omega) = \frac{F_\ba(x_p, y_p, \omega)}{k^r_n+k_\ba^r - m \omega^2 +i(k^i_n+k_\ba^i)}\phi_n(y_p)
\end{equation}
The power spectrum density of $\xi^\mathrm{SN}(y_p, \omega)$ is strongly peaked around the resonance angular frequency $\omega_n=\sqrt{(k^r_n+k_\ba^r)/m}$. We can therefore neglect the frequency dependence of other terms when computing the mean square deflection $\langle \xi_n^\mathrm{SN}(y_p)^2 \rangle$ of mode $n$ measured in $y_p$ by integrating the power spectrum on all frequencies around $\omega_n$, which leads to:
\begin{align}
\!\!\langle \xi_n^\mathrm{SN}(y_p)^2 \rangle &= \frac{\pi\omega_n\phi_n^2(y_p)}{2(k^r_n+k_\ba^r)(k^i_r+k_\ba^i)}S_{F_\ba}(x_p, y_p, \omega_n)\!\\
&= \frac{\pi\omega_n}{(k^r_n+k_\ba^r)(k^i_n+k_\ba^i)}|k_\ba|^2\frac{\abar^2}{a'(x_p)^2} \frac{\epsilon}{\Pbar}.
\end{align}
 The variance of the quantum back-action-induced motion detected at the probe position $y_p$ is therefore essentially determined by the profile of the complex stiffness back-action $|k_\ba|$. In particular, if the modification of the mechanical resonance is weak (no sizeable frequency shift or damping alteration), the shot noise driven deflection noise follows the square modulus of $k_\ba$, presented in Fig.~\ref{Figk_ba}. In this case, the motion standard deviation is proportional to $\Pbar$ (since $k_\ba\propto\Pbar$). Conversely, if $(k^i_n+k_\ba^i)$ significantly varies, the power scaling of the measured fluctuations may become challenging to analyze, with a net motion variance that may either increase or stagnate, depending on the thermal back-action gradient sign. Thereby, it is legitimate to question whether the observation of such back-action-driven vibrations enables to conclude as to the quantum origin of the underlying driving processes, as opposed to fluctuations resulting from thermal agitation, to which the next section is dedicated.

\section{Thermal back-action noise} \label{section:thermalnoise}

In this section, we tackle the question of the thermal noise of the resonator, a nontrivial problem since the hypotheses of the fluctuation-dissipation theorem are not fulfilled. Indeed, the system is in a non equilibrium steady state (NESS), as it bears a constant heat flow from the probe beam position $y_p$ to the thermostat, here the clamp assumed to be at a fixed temperature $T_0$. The temperature itself is thus a field, and the fluctuation-dissipation theorem needs to be extended to this NESS. We follow Refs.~\onlinecite{Geitner2017, Komori2018, Fontana2020, Fontana2023}, which demonstrate that the amplitude of the thermal noise in this instance is
\begin{equation} \label{eq.ThNoiseNess}
\langle \xi_n^\mathrm{TN}(y_p)^2 \rangle = \frac{k_B T^*}{k^r_n+k_\ba^r}\frac{k_n^i}{k^i_n+k_\ba^i} \phi_n(y_p)^2, 
\end{equation}
where the superscript $^\mathrm{TN}$ stands for thermal noise driven, and $T^*$ is the equivalent temperature of the oscillator, defined by
\begin{align}
T^* &=  \frac{1}{k_n^i}\int_{0}^{L} \Tbar(y) E^iI \phi_n''(y)^2 \dd y \\
& =  T_0+ \bar\theta \frac{1}{Lk_n^i}\int_{0}^{L} \min(y, y_p) E^iI \phi_n''(y)^2 \dd y.
\end{align}
$T^*$ describes the amplitude of the thermal noise driven fluctuations, and corresponds to the average temperature of the system with a weighing that matches the normalized local dissipation of mechanical energy $E^iI \phi_n''(y)^2/k_n^i$. If $\bar\theta\ll T_0$ (small heating effects), then $T^* \sim T_0$ and Eq.~\ref{eq.ThNoiseNess} is very similar to the equipartition at equilibrium, except that the effective temperature is $T_0 k^i_n/(k^i_n+k_\ba^i)$. Once again, we see the effect of the back-action that effectively cools or warms the resonant mode depending on the sign of $k_\ba^i$.

\begin{figure}[tb]
\begin{center}
\includegraphics{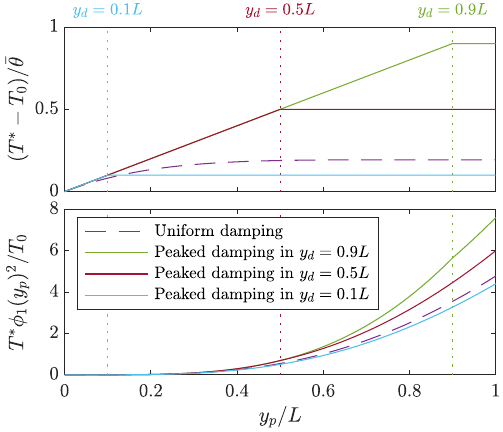}
\caption{Thermal noise driven fluctuations. (a) The NESS contribution to the thermal noise is summarized in $T^*$, the effective temperature that should be taken into account when extending the fluctuation-dissipation theorem to this non-equilibrium steady state. The overhead to $T_0$ is proportional to $\bar \theta$, and increases with the probe position $y_p$. Four hypotheses on the mechanical dissipation distribution are illustrated here: uniform damping (purple dashed line, computed for the first normal mode), localized damping in $y_d=0.1, 0.5$ or $0.9$. (b) Appart from the amplification/damping refactor coming from the back-action stiffness, the thermal noise profile when scanning the probe along the resonator is proportional to $T^* \phi_n(y_p)^2$. It is illustrated here for $\bar \theta=T_0$ and $n=1$, and mainly follows the square of the mode shape since the effective temperature is a pre-factor bounded between 1 and 2 in this case.}
\label{Fig.Tstar}
\end{center}
\end{figure}

The NESS contribution to $T^*$ comes from the non-uniform temperature field, $\bar \theta \min(y, y_p)/L$, proportional to $\Pbar$. It depends on the probe position and power, due to the interplay between the fields of temperature and dissipation. The latter is expressed here with the imaginary part of the Young modulus and can have several origins, such as distributed thermoelastic or anelastic damping, internal friction~\cite{Saulson1990, Geitner2017, Fontana2023}, or be point-like and localized at the defects' positions or at the clamp~\cite{Geitner2017, Fontana2020}. A reasonable assumption is that the spatial distributions of this dissipation and of the linear thermal bending coefficient are correlated: a uniform $\beta$ corresponds to a uniform $E^i$, a peaked $\beta$ in $y_d$ leads to a localized damping $E^i\propto E^i_d \delta^D(y-y_d)$. We report in Fig.~\ref{Fig.Tstar} the dependency of $T^*$ as a function of $y_p$ under those various hypotheses, and the corresponding thermal noise driven fluctuations profile. The latter is increasing when the probe gets closer to the free end of the resonator, and the dissipation distribution only has a mild influence on this qualitative shape.

At this stage, we can conclude that the shot and thermal noises are distinguished by their evolution as a function of $y_p$, while showing the same linear power dependence. It is however not possible to draw general conclusions so as to the relative weights of these two effects, which stems from the thermal bending parameter $\beta$, which essentially determines the quantum back-action. This parameter may acquire sizeable values (corresponding to large bending under small temperature variations), with a correspondingly large shot noise  therefore having to be accounted for, together with the NESS contribution. In the next section, we further consider the consequences of the combination of these two noises, in particular in the context of force sensing.

\section{Back-action noises in force sensing} \label{section:forcenoise}

\begin{figure}[htbp]
\begin{center}
\includegraphics{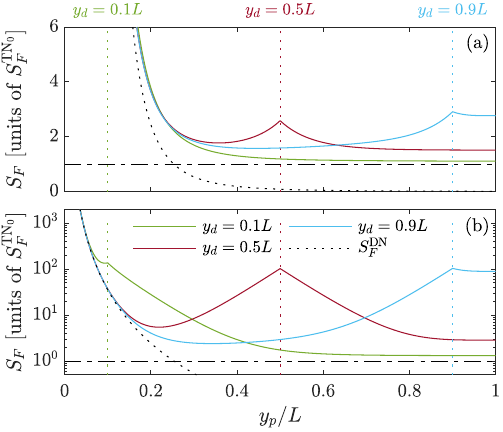}
\caption{Total force noise in a photothermal back-action-driven mechanical resonator computed from Eq.~\ref{eq.SF}, expressed in unit of equilibrium thermal noise $S_F^\mathrm{TN_0}=2 k_n^i k_BT_0/\omega_n$. The plain lines are computed for $|\alpha|L=10$ and a single defect contribution to $\beta$ and $E^i$, localized in $y_d=0.1, 0.5$ or $0.9$. The dashed-dotted line stands for the equilibrium thermal limit $S_F^\mathrm{TN_0}$, and the dotted line to the detection noise contribution. (a) Case where the three first contributions are assumed to have the same, maximal magnitude ($\mathrm{max}(S_F^\mathrm{TN^*})=\mathrm{max}(S_F^\mathrm{SN})=S_F^\mathrm{TN_0}$). (b) Strongly driven, quantum dominated case: the maximal magnitudes of the NESS term is $\mathrm{max}(S_F^\mathrm{TN^*})=3\times S_F^\mathrm{TN_0}$ and the one of the shot noise contribution is $\mathrm{max}(S_F^\mathrm{SN})=100\times S_F^\mathrm{TN_0}$.}
\label{Fig.sensitivity}
\end{center}
\end{figure}

Here we assume that force sensing is performed from a local measurement of the deformation of the interacting resonator. The estimated force is therefore deduced using Eq.~\ref{eq.OHba}, which we rewrite here as:
\begin{equation}
    F_n=\chi(x_p,y_p,\omega_n)\frac{\xi(y_p,\omega_n)}{\phi_n(y_p)},
\end{equation}
where $F_n$ is the projection of the external force field to be measured on the considered mode $\phi_n$, and $\chi=k^r_n+k_\ba^r - m \omega^2 +i(k^i_n+k_\ba^i)$ is the harmonic oscillator response function. On top of the thermal ($\xi^\mathrm{TN}$) and shot ($\xi^\mathrm{SN}$) noises described previously, we add the detection noise ($\xi^\mathrm{DN}$) for which we assume a white noise spectrum. The total equivalent force noise detected in presence of the back-action processes is then:
\begin{equation} \label{eq.SF}
    S_F = S_F^\mathrm{TN_0} + S_F^\mathrm{TN^*} + S_F^\mathrm{SN} + S_F^\mathrm{DN},
\end{equation}
with $S_F^\mathrm{TN_0}(\omega)=2 k_n^i k_BT_0/\omega_n$ the thermal noise associated with the thermal bath (that is, in absence of measurement-induced heating), $S_F^\mathrm{TN^*}(x_p, y_p, \omega)=2 k_n^i k_B(T^*-T_0)/\omega_n$ the NESS contribution associated with the heat flow along the NW, $S_F^\mathrm{SN}(x_p, y_p, \omega)$ the force spectral density associated with the shot noise back-action, and $S_F^\mathrm{DN}(y_p, \omega))\propto 1/\phi_n(y_p)^2$ the detection noise floor. The latter term diverges on nodes of the mode shape, and in particular at the origin: though the NESS and shot noise contributions are simultaneously minimal for $y_p=0$, it make no sense measuring there as the sensitivity also vanishes. $S_F$ is plotted in Fig.~\ref{Fig.sensitivity} for equivalent contributions from all three driving sources (that is, $\mathrm{max}(S_F^\mathrm{TN^*})=\mathrm{max}(S_F^\mathrm{SN})=S_F^\mathrm{TN_0}$, top panel) and for dominating quantum back-action effects ($\mathrm{max}(S_F^\mathrm{TN^*})=3\times S_F^\mathrm{TN_0}$, $\mathrm{max}(S_F^\mathrm{SN})=100\times S_F^\mathrm{TN_0}$, bottom panel). It shows that measuring the mechanical motion at the tip of the NW yields a close to minimal force noise in presence of defects implanted far from the tip ($y_d/L=0.1$ and $y_d/L=0.5$). However and interestingly, this is no longer true for a defect close to the apex ($y_d=0.9$): in this case, the measurement becomes substantially noisier, due to the contribution of the quantum back-action. By moving closer to the clamp, one sees that significant reductions of the force noise may be achieved. This may be particularly noteworthy in the context of nanomechanical scanning probes \cite{pigeau2015observation,budakian2024roadmap}, which generally functionalize the tip of a free-standing micro-/nanomechanical resonator, and measure its mechanical response close to it, therefore with potential important quantum back-action noise influence.

\section{Application cases} \label{section:appcase}

The experimental relevance of our approach can be confirmed by considering the case of cylindrical SiC or Si nano-optomechanical scanning probes, which have recently been praised for their high sensitivity and reliability \cite{sanii2010high,arcizet2011single,sahafi2019ultralow,fogliano2021ultrasensitive}. Assuming typical length $L=\SI{100}{\mu m}$ and diameter $2R=\SI{100}{nm}$, and with the following approximate material values $E\simeq \SI{300}{GPa}$, $\rho\simeq \SI{3000}{kg/m^3}$ (yielding a mechanical resonance frequency $\omega_1/2\pi\simeq \SI{14}{kHz}$), $c\simeq \SI{750}{J.K^{-1}.kg^{-1}}$, $\kappa\simeq\SI{10}{W.K^{-1}.m^{-1}}$ \cite{Nigues2014a,valentin2013comprehensive}, one gets $|\alpha|L\simeq 14$, similar to the intermediate case treated here (middle column of Fig.~\ref{Figk_ba}, Fig.~\ref{Fig.sensitivity}). Assuming that the optomechanical readout is achieved with an optical power $P=\SI{100}{\mu W}$, in presence of an average absorption rate of $\overline{a}=0.1\%$, and that the absorption gradient is essentially set by the optical waist $w_0=\SI{1}{\mu m}$, i.e. $\dd a/\dd x_p\simeq a/w_0 = \SI{E3}{m^{-1}}$, we find $\mathrm{max}\left(|k_{\mathrm{ba}}|\right)\simeq \SI{2.5e-9}{N/m}$ for a point-like, centered defect with $y_{\mathrm{d}}/L=0.5$ and $\beta_d=\SI{e3}{K^{-1}}$. Note that this latter value means that, when hitting the defect with the laser probe, the nanowire bends so that its tip laterally moves approximately from $\SI{2}{nm}$, that is, only $2\%$ of the NW diameter, while experimental observations routinely range within several diameters, suggesting that standard values of $\beta_d$ are likely to be orders of magnitude larger. Gauging upon the efficiency of the back-action stiffness can be done by comparing it to the dissipation, $|k_{\mathrm{ba}}/k_n^i|\simeq Q|k_{\mathrm{ba}}/m_{\mathrm{eff}}\omega_1^2|\simeq 6$ ($m_{\mathrm{eff}}\simeq m/4$ the effective mass, and $Q\simeq\num{E4}$ the quality factor): this means that such measurement would operate with an absorbed probe power that is more than five times the parametric instability threshold. The prominence of the dynamical back-action would only be reinforced at higher Q-factor, an issue that is ubiquitous to the development of ultra-low dissipation mechanical systems \cite{ghadimi2018elastic, ren2020two}.

Another type of systems specifically prone to this phenomenology are ultra-low diameter, high aspect-ratio nanomechanical systems, including carbon nanotubes and semiconducting NWs \cite{postma2005dynamic, moser2014b, tavernarakis2018optomechanics, pairis2019shot}. On top of a generally enhanced sensitivity to defects, stemming from their extremely reduced size, these systems generally are amongst those with the lowest thermal noise, which results in a relative reinforcement of their responsivity towards external fluctuation processes, including measurement quantum back-action. Since those effects are parameterized by $\beta_d^2$, they may acquire very large spectral densities, to the point that they overwhelmingly dominate other noise sources. This, together with a full, quantitative verification of the hereby reported theoretical model, is the object of a recent work from our collaboration \cite{chardin2024}, and matches the choice of parameters used in Fig.~\ref{Fig.sensitivity}(b).

\section{Conclusions and perspectives} \label{sec:ccl}

In conclusion, measuring the vibrations of micro- and nanoresonators using a high-energy probe beam will unavoidably result in heating those tiny devices. The induced temperature field feeds back to the motion of the resonator, yielding changes in its vibrational response (frequency, amplitude, and quality factor). This back-action effect strongly depends on the probe position, the thermal diffusion length, and the spatial distribution of the thermomechanical coupling mechanism. Assuming a strong correlation between this thermomechanical coupling ($\beta$) and the mechanical losses ($E^i$), we are able to quantitatively describe the fluctuational processes associated with this back-action, including the thermal noise of the system in the nonequilibrium steady state generated by the heat flow, and the quantum noise due to the shot noise of the probe. The signatures of all these processes \emph{when scanning the probe} along and across the resonator are found to be very specific, especially when the losses / thermo-mechanical coupling are strongly localized, which is a commonly expected case. For functionalized sensors for example, the modification of the resonator by e.g. grafted molecules~\cite{Lavrik2004, Garcia2002}, plasmonic coatings \cite{thijssen2013plasmon}, strain-coupled single quantum emitters~\cite{montinaro2014quantum,yeo2014strain,teissier2014strain}, magnetic nanoparticles~\cite{rugar2004single,arcizet2011single,budakian2024roadmap}, purposely alters the mechanical response, with predictable side effects in terms of backaction and associated noise. For defect-free resonators, the clamp itself is a singular area where broken symmetries or peculiarities~\cite{kettler2021inducing}, together with clamping losses~\cite{Hao2003,rieger2014energy, Geitner2017, Fontana2020, Fontana2023}, are rather the rule than the exception. The analysis of the real (frequency shift) and imaginary (damping) part of the response function and of the noise profile provides a rich tool to characterize the nanoresonator: a single model should encompass all three observations, making the conclusions very robust. Our approach is certainly applicable in many instances and should be of general and wide interest.

The analytical framework presented in this article is exemplified with the study of a 1D resonator, with clamped-free boundary conditions, point like probe beam, all figures being generated considering the first resonant mode. Exploring the signatures with other modes simply requires changing the strain profile to the corresponding eigenfunction $\phi_n(y)$. Experimentally, this should prove to be very informative, as the thermal diffusion length changes significantly with frequency, while the spatial distribution of the thermomechanical coupling remains essentially unaltered. In particular, such a multimode approach may provide additional accuracy in addressing the nature and location of mechanical losses.

We limited the study to probe beams of small spatial extension compared to $L$ and to the thermal diffusion length $1/|\alpha|$. As mentioned in section \ref{section:heat}, our solutions are however Green functions in the linear framework we describe. An additional convolution step would be necessary to deal with distributed sensing, for example to model the experiments of Refs.~\onlinecite{Morell-2019, Iadanza-2020}. Our approach remains valid in general as soon as part of the incoming probe energy is dissipated by the measured mechanical device, which virtually applies to all transduction schemes (on-chip electrodes~\cite{unterreithmeier2009universal} for example).

Extending our approach to various boundary conditions is straightforward for 1D oscillators. For example, for double clamp bridges, Eq.~\ref{eqBCthL} changes to fixing the temperature in $L$: $T(L)=T_0$. The static and dynamic thermal profiles can subsequently be solved using the same methodology as that outlined in Section \ref{section:heat}, with the mechanical modes being also analytically computable. From there, hypotheses on $\beta$ (and $E^i$) can be made to test different scenarios, and compared to experimental data, thus providing quantitative insights about the dominant loss mechanisms, their spatial location and possible mitigation strategies. For 2D or 3D resonators, the methodology remains valid but can turn very complex, as the analytical description of the temperature and deformation fields would most likely be limited to specific geometries and boundary conditions. In such cases, turning to numerical simulations may become unavoidable~\cite{primo2021accurate}, making it a less efficient tool to probe a vast range of thermo-mechanical coupling hypotheses or probe positions.

Another possible extension of this work is towards pump-probe approaches. Indeed, one key learning of this study is that a wealth of information can be unraveled by scanning the probe beam along the resonator. In our framework, this single beam acts both as a probe (measuring the deflection) and as a pump (creating the temperature field that interacts with the motion). An even richer behavior could be explored by decoupling the two functions, i.e. using a high-power pump to create the thermomechanical effects, and a much weaker one for probing the response at a different location~\cite{khivrich2019nanomechanical, Gouriou2023PhDThesis}, for instance where the sensitivity is maximized.

Lastly, our analytical derivation shows that, by controlling the distribution of $\beta$, both the thermomechanical coupling and the back-action dynamics can actually be tuned with the position of the probe along the resonator. It should then be possible to engineer the back-action so as to selectively address the damping/effective temperature with no associated frequency detuning, or vice versa, to control the frequency with no corresponding modification of the dissipation. Such a perspective could even be expanded with a pump-probe configuration, opening the perspective of spatially resolved, non-equilibrium thermal state engineering. 

\acknowledgments

This work has been partially funded by projects ANR-16-CE09-0010, ANR-22-CE42-0022 and ERC StG 758794. We thank J. Claudon, M. Hocevar, A. Fontana, and O. Bourgeois for constructive scientific discussions.

\bibliography{bib}

\begin{thebibliography}{64}%
\makeatletter
\providecommand \@ifxundefined [1]{%
 \@ifx{#1\undefined}
}%
\providecommand \@ifnum [1]{%
 \ifnum #1\expandafter \@firstoftwo
 \else \expandafter \@secondoftwo
 \fi
}%
\providecommand \@ifx [1]{%
 \ifx #1\expandafter \@firstoftwo
 \else \expandafter \@secondoftwo
 \fi
}%
\providecommand \natexlab [1]{#1}%
\providecommand \enquote  [1]{``#1''}%
\providecommand \bibnamefont  [1]{#1}%
\providecommand \bibfnamefont [1]{#1}%
\providecommand \citenamefont [1]{#1}%
\providecommand \href@noop [0]{\@secondoftwo}%
\providecommand \href [0]{\begingroup \@sanitize@url \@href}%
\providecommand \@href[1]{\@@startlink{#1}\@@href}%
\providecommand \@@href[1]{\endgroup#1\@@endlink}%
\providecommand \@sanitize@url [0]{\catcode `\\12\catcode `\$12\catcode
  `\&12\catcode `\#12\catcode `\^12\catcode `\_12\catcode `\%12\relax}%
\providecommand \@@startlink[1]{}%
\providecommand \@@endlink[0]{}%
\providecommand \url  [0]{\begingroup\@sanitize@url \@url }%
\providecommand \@url [1]{\endgroup\@href {#1}{\urlprefix }}%
\providecommand \urlprefix  [0]{URL }%
\providecommand \Eprint [0]{\href }%
\providecommand \doibase [0]{https://doi.org/}%
\providecommand \selectlanguage [0]{\@gobble}%
\providecommand \bibinfo  [0]{\@secondoftwo}%
\providecommand \bibfield  [0]{\@secondoftwo}%
\providecommand \translation [1]{[#1]}%
\providecommand \BibitemOpen [0]{}%
\providecommand \bibitemStop [0]{}%
\providecommand \bibitemNoStop [0]{.\EOS\space}%
\providecommand \EOS [0]{\spacefactor3000\relax}%
\providecommand \BibitemShut  [1]{\csname bibitem#1\endcsname}%
\let\auto@bib@innerbib\@empty
\bibitem [{\citenamefont {O'Connell}\ \emph {et~al.}(2010)\citenamefont
  {O'Connell}, \citenamefont {Hofheinz}, \citenamefont {Ansmann}, \citenamefont
  {Bialczak}, \citenamefont {Lenander}, \citenamefont {Lucero}, \citenamefont
  {Neeley}, \citenamefont {Sank}, \citenamefont {Wang}, \citenamefont {Weides}
  \emph {et~al.}}]{o2010quantum}%
  \BibitemOpen
  \bibfield  {author} {\bibinfo {author} {\bibfnamefont {A.~D.}\ \bibnamefont
  {O'Connell}}, \bibinfo {author} {\bibfnamefont {M.}~\bibnamefont {Hofheinz}},
  \bibinfo {author} {\bibfnamefont {M.}~\bibnamefont {Ansmann}}, \bibinfo
  {author} {\bibfnamefont {R.~C.}\ \bibnamefont {Bialczak}}, \bibinfo {author}
  {\bibfnamefont {M.}~\bibnamefont {Lenander}}, \bibinfo {author}
  {\bibfnamefont {E.}~\bibnamefont {Lucero}}, \bibinfo {author} {\bibfnamefont
  {M.}~\bibnamefont {Neeley}}, \bibinfo {author} {\bibfnamefont
  {D.}~\bibnamefont {Sank}}, \bibinfo {author} {\bibfnamefont {H.}~\bibnamefont
  {Wang}}, \bibinfo {author} {\bibfnamefont {M.}~\bibnamefont {Weides}}, \emph
  {et~al.},\ }\bibfield  {title} {\bibinfo {title} {Quantum ground state and
  single-phonon control of a mechanical resonator},\ }\href
  {https://doi.org/10.1038/nature08967} {\bibfield  {journal} {\bibinfo
  {journal} {Nature}\ }\textbf {\bibinfo {volume} {464}},\ \bibinfo {pages}
  {697} (\bibinfo {year} {2010})}\BibitemShut {NoStop}%
\bibitem [{\citenamefont {Gil-Santos}\ \emph {et~al.}(2010)\citenamefont
  {Gil-Santos}, \citenamefont {Ramos}, \citenamefont {Mart\'{\i}nez},
  \citenamefont {Fern\'{a}ndez-Reg\'{u}lez}, \citenamefont {Garc\'{\i}a},
  \citenamefont {{San Paulo}}, \citenamefont {Calleja},\ and\ \citenamefont
  {Tamayo}}]{Gil-Santos2010}%
  \BibitemOpen
  \bibfield  {author} {\bibinfo {author} {\bibfnamefont {E.}~\bibnamefont
  {Gil-Santos}}, \bibinfo {author} {\bibfnamefont {D.}~\bibnamefont {Ramos}},
  \bibinfo {author} {\bibfnamefont {J.}~\bibnamefont {Mart\'{\i}nez}}, \bibinfo
  {author} {\bibfnamefont {M.}~\bibnamefont {Fern\'{a}ndez-Reg\'{u}lez}},
  \bibinfo {author} {\bibfnamefont {R.}~\bibnamefont {Garc\'{\i}a}}, \bibinfo
  {author} {\bibfnamefont {A.}~\bibnamefont {{San Paulo}}}, \bibinfo {author}
  {\bibfnamefont {M.}~\bibnamefont {Calleja}},\ and\ \bibinfo {author}
  {\bibfnamefont {J.}~\bibnamefont {Tamayo}},\ }\bibfield  {title} {\bibinfo
  {title} {{Nanomechanical mass sensing and stiffness spectrometry based on
  two-dimensional vibrations of resonant nanowires.}},\ }\href
  {https://doi.org/10.1038/nnano.2010.151} {\bibfield  {journal} {\bibinfo
  {journal} {Nat. Nanotech.}\ }\textbf {\bibinfo {volume} {5}},\ \bibinfo
  {pages} {641} (\bibinfo {year} {2010})}\BibitemShut {NoStop}%
\bibitem [{\citenamefont {Chan}\ \emph {et~al.}(2011)\citenamefont {Chan},
  \citenamefont {Alegre}, \citenamefont {Safavi-Naeini}, \citenamefont {Hill},
  \citenamefont {Krause}, \citenamefont {Gr{\"o}blacher}, \citenamefont
  {Aspelmeyer},\ and\ \citenamefont {Painter}}]{chan2011laser}%
  \BibitemOpen
  \bibfield  {author} {\bibinfo {author} {\bibfnamefont {J.}~\bibnamefont
  {Chan}}, \bibinfo {author} {\bibfnamefont {T.~M.}\ \bibnamefont {Alegre}},
  \bibinfo {author} {\bibfnamefont {A.~H.}\ \bibnamefont {Safavi-Naeini}},
  \bibinfo {author} {\bibfnamefont {J.~T.}\ \bibnamefont {Hill}}, \bibinfo
  {author} {\bibfnamefont {A.}~\bibnamefont {Krause}}, \bibinfo {author}
  {\bibfnamefont {S.}~\bibnamefont {Gr{\"o}blacher}}, \bibinfo {author}
  {\bibfnamefont {M.}~\bibnamefont {Aspelmeyer}},\ and\ \bibinfo {author}
  {\bibfnamefont {O.}~\bibnamefont {Painter}},\ }\bibfield  {title} {\bibinfo
  {title} {Laser cooling of a nanomechanical oscillator into its quantum ground
  state},\ }\href {https://doi.org/10.1038/nature10461} {\bibfield  {journal}
  {\bibinfo  {journal} {Nature}\ }\textbf {\bibinfo {volume} {478}},\ \bibinfo
  {pages} {89} (\bibinfo {year} {2011})}\BibitemShut {NoStop}%
\bibitem [{\citenamefont {Liu}\ \emph {et~al.}(2021)\citenamefont {Liu},
  \citenamefont {Liu}, \citenamefont {Ren}, \citenamefont {Ma}, \citenamefont
  {Dong}, \citenamefont {Zhou},\ and\ \citenamefont {Lee}}]{Liu-2021}%
  \BibitemOpen
  \bibfield  {author} {\bibinfo {author} {\bibfnamefont {X.}~\bibnamefont
  {Liu}}, \bibinfo {author} {\bibfnamefont {W.}~\bibnamefont {Liu}}, \bibinfo
  {author} {\bibfnamefont {Z.}~\bibnamefont {Ren}}, \bibinfo {author}
  {\bibfnamefont {Y.}~\bibnamefont {Ma}}, \bibinfo {author} {\bibfnamefont
  {B.}~\bibnamefont {Dong}}, \bibinfo {author} {\bibfnamefont {G.}~\bibnamefont
  {Zhou}},\ and\ \bibinfo {author} {\bibfnamefont {C.}~\bibnamefont {Lee}},\
  }\bibfield  {title} {\bibinfo {title} {Progress of optomechanical micro/nano
  sensors: a review},\ }\href {https://doi.org/10.1080/15599612.2021.1986612}
  {\bibfield  {journal} {\bibinfo  {journal} {Int. J. Optomechatronics}\
  }\textbf {\bibinfo {volume} {15}},\ \bibinfo {pages} {120} (\bibinfo {year}
  {2021})}\BibitemShut {NoStop}%
\bibitem [{\citenamefont {Aspelmeyer}\ \emph {et~al.}(2014)\citenamefont
  {Aspelmeyer}, \citenamefont {Kippenberg},\ and\ \citenamefont
  {Marquardt}}]{aspelmeyer2014cavity}%
  \BibitemOpen
  \bibfield  {author} {\bibinfo {author} {\bibfnamefont {M.}~\bibnamefont
  {Aspelmeyer}}, \bibinfo {author} {\bibfnamefont {T.~J.}\ \bibnamefont
  {Kippenberg}},\ and\ \bibinfo {author} {\bibfnamefont {F.}~\bibnamefont
  {Marquardt}},\ }\bibfield  {title} {\bibinfo {title} {Cavity optomechanics},\
  }\href {https://doi.org/10.1103/RevModPhys.86.1391} {\bibfield  {journal}
  {\bibinfo  {journal} {Rev. Mod. Phys.}\ }\textbf {\bibinfo {volume} {86}},\
  \bibinfo {pages} {1391} (\bibinfo {year} {2014})}\BibitemShut {NoStop}%
\bibitem [{\citenamefont {Saulson}(1990)}]{Saulson1990}%
  \BibitemOpen
  \bibfield  {author} {\bibinfo {author} {\bibfnamefont {P.~R.}\ \bibnamefont
  {Saulson}},\ }\bibfield  {title} {\bibinfo {title} {Thermal noise in
  mechanical experiments},\ }\href {https://doi.org/10.1103/PhysRevD.42.2437}
  {\bibfield  {journal} {\bibinfo  {journal} {Phys. Rev. D}\ }\textbf {\bibinfo
  {volume} {42}},\ \bibinfo {pages} {2437} (\bibinfo {year}
  {1990})}\BibitemShut {NoStop}%
\bibitem [{\citenamefont {Blaikie}\ \emph {et~al.}(2019)\citenamefont
  {Blaikie}, \citenamefont {Miller},\ and\ \citenamefont
  {Alem{\'a}n}}]{blaikie2019fast}%
  \BibitemOpen
  \bibfield  {author} {\bibinfo {author} {\bibfnamefont {A.}~\bibnamefont
  {Blaikie}}, \bibinfo {author} {\bibfnamefont {D.}~\bibnamefont {Miller}},\
  and\ \bibinfo {author} {\bibfnamefont {B.~J.}\ \bibnamefont {Alem{\'a}n}},\
  }\bibfield  {title} {\bibinfo {title} {A fast and sensitive room-temperature
  graphene nanomechanical bolometer},\ }\href
  {https://doi.org/10.1038/s41467-019-12562-2} {\bibfield  {journal} {\bibinfo
  {journal} {Nat. Commun.}\ }\textbf {\bibinfo {volume} {10}},\ \bibinfo
  {pages} {4726} (\bibinfo {year} {2019})}\BibitemShut {NoStop}%
\bibitem [{\citenamefont {Chen}\ \emph {et~al.}(2022)\citenamefont {Chen},
  \citenamefont {Li},\ and\ \citenamefont {Zheng}}]{Chen-2022}%
  \BibitemOpen
  \bibfield  {author} {\bibinfo {author} {\bibfnamefont {Z.}~\bibnamefont
  {Chen}}, \bibinfo {author} {\bibfnamefont {J.}~\bibnamefont {Li}},\ and\
  \bibinfo {author} {\bibfnamefont {Y.}~\bibnamefont {Zheng}},\ }\bibfield
  {title} {\bibinfo {title} {Heat-mediated optical manipulation},\ }\href
  {https://doi.org/10.1021/acs.chemrev.1c00626} {\bibfield  {journal} {\bibinfo
   {journal} {Chem. Rev.}\ }\textbf {\bibinfo {volume} {122}},\ \bibinfo
  {pages} {3122} (\bibinfo {year} {2022})}\BibitemShut {NoStop}%
\bibitem [{\citenamefont {Braginski{\u \i}}\ and\ \citenamefont
  {Manukin}(1967)}]{braginski1967ponderomotive}%
  \BibitemOpen
  \bibfield  {author} {\bibinfo {author} {\bibfnamefont {V.~B.}\ \bibnamefont
  {Braginski{\u \i}}}\ and\ \bibinfo {author} {\bibfnamefont {A.~B.}\
  \bibnamefont {Manukin}},\ }\bibfield  {title} {\bibinfo {title}
  {Ponderomotive effects of electromagnetic radiation},\ }\href
  {http://jetp.ras.ru/cgi-bin/e/index/e/25/4/p653?a=list} {\bibfield  {journal}
  {\bibinfo  {journal} {Sov. Phys. JETP}\ }\textbf {\bibinfo {volume} {25}},\
  \bibinfo {pages} {653} (\bibinfo {year} {1967})}\BibitemShut {NoStop}%
\bibitem [{\citenamefont {Metzger}\ and\ \citenamefont
  {Karrai}(2004)}]{metzger2004cavity}%
  \BibitemOpen
  \bibfield  {author} {\bibinfo {author} {\bibfnamefont {C.~H.}\ \bibnamefont
  {Metzger}}\ and\ \bibinfo {author} {\bibfnamefont {K.}~\bibnamefont
  {Karrai}},\ }\bibfield  {title} {\bibinfo {title} {Cavity cooling of a
  microlever},\ }\href {https://doi.org/10.1038/nature03118} {\bibfield
  {journal} {\bibinfo  {journal} {Nature}\ }\textbf {\bibinfo {volume} {432}},\
  \bibinfo {pages} {1002} (\bibinfo {year} {2004})}\BibitemShut {NoStop}%
\bibitem [{\citenamefont {Verhagen}\ \emph {et~al.}(2012)\citenamefont
  {Verhagen}, \citenamefont {Del{\'e}glise}, \citenamefont {Weis},
  \citenamefont {Schliesser},\ and\ \citenamefont
  {Kippenberg}}]{verhagen2012quantum}%
  \BibitemOpen
  \bibfield  {author} {\bibinfo {author} {\bibfnamefont {E.}~\bibnamefont
  {Verhagen}}, \bibinfo {author} {\bibfnamefont {S.}~\bibnamefont
  {Del{\'e}glise}}, \bibinfo {author} {\bibfnamefont {S.}~\bibnamefont {Weis}},
  \bibinfo {author} {\bibfnamefont {A.}~\bibnamefont {Schliesser}},\ and\
  \bibinfo {author} {\bibfnamefont {T.~J.}\ \bibnamefont {Kippenberg}},\
  }\bibfield  {title} {\bibinfo {title} {Quantum-coherent coupling of a
  mechanical oscillator to an optical cavity mode},\ }\href
  {https://doi.org/10.1038/nature10787} {\bibfield  {journal} {\bibinfo
  {journal} {Nature}\ }\textbf {\bibinfo {volume} {482}},\ \bibinfo {pages}
  {63} (\bibinfo {year} {2012})}\BibitemShut {NoStop}%
\bibitem [{\citenamefont {Nigu{\`e}s}\ \emph {et~al.}(2015)\citenamefont
  {Nigu{\`e}s}, \citenamefont {Siria},\ and\ \citenamefont
  {Verlot}}]{Nigues2014a}%
  \BibitemOpen
  \bibfield  {author} {\bibinfo {author} {\bibfnamefont {A.}~\bibnamefont
  {Nigu{\`e}s}}, \bibinfo {author} {\bibfnamefont {A.}~\bibnamefont {Siria}},\
  and\ \bibinfo {author} {\bibfnamefont {P.}~\bibnamefont {Verlot}},\
  }\bibfield  {title} {\bibinfo {title} {Dynamical backaction cooling with free
  electrons},\ }\href {https://doi.org/10.1038/ncomms9104} {\bibfield
  {journal} {\bibinfo  {journal} {Nat. Commun.}\ }\textbf {\bibinfo {volume}
  {6}},\ \bibinfo {pages} {8104} (\bibinfo {year} {2015})}\BibitemShut
  {NoStop}%
\bibitem [{\citenamefont {Metzger}\ \emph {et~al.}(2008)\citenamefont
  {Metzger}, \citenamefont {Ludwig}, \citenamefont {Neuenhahn}, \citenamefont
  {Ortlieb}, \citenamefont {Favero}, \citenamefont {Karrai},\ and\
  \citenamefont {Marquardt}}]{Metzger2008}%
  \BibitemOpen
  \bibfield  {author} {\bibinfo {author} {\bibfnamefont {C.}~\bibnamefont
  {Metzger}}, \bibinfo {author} {\bibfnamefont {M.}~\bibnamefont {Ludwig}},
  \bibinfo {author} {\bibfnamefont {C.}~\bibnamefont {Neuenhahn}}, \bibinfo
  {author} {\bibfnamefont {A.}~\bibnamefont {Ortlieb}}, \bibinfo {author}
  {\bibfnamefont {I.}~\bibnamefont {Favero}}, \bibinfo {author} {\bibfnamefont
  {K.}~\bibnamefont {Karrai}},\ and\ \bibinfo {author} {\bibfnamefont
  {F.}~\bibnamefont {Marquardt}},\ }\bibfield  {title} {\bibinfo {title}
  {Self-induced oscillations in an optomechanical system driven by bolometric
  backaction},\ }\href {https://doi.org/10.1103/PhysRevLett.101.133903}
  {\bibfield  {journal} {\bibinfo  {journal} {Phys. Rev. Lett.}\ }\textbf
  {\bibinfo {volume} {101}},\ \bibinfo {pages} {133903} (\bibinfo {year}
  {2008})}\BibitemShut {NoStop}%
\bibitem [{\citenamefont {Tavernarakis}\ \emph {et~al.}(2018)\citenamefont
  {Tavernarakis}, \citenamefont {Stavrinadis}, \citenamefont {Nowak},
  \citenamefont {Tsioutsios}, \citenamefont {Bachtold},\ and\ \citenamefont
  {Verlot}}]{tavernarakis2018optomechanics}%
  \BibitemOpen
  \bibfield  {author} {\bibinfo {author} {\bibfnamefont {A.}~\bibnamefont
  {Tavernarakis}}, \bibinfo {author} {\bibfnamefont {A.}~\bibnamefont
  {Stavrinadis}}, \bibinfo {author} {\bibfnamefont {A.}~\bibnamefont {Nowak}},
  \bibinfo {author} {\bibfnamefont {I.}~\bibnamefont {Tsioutsios}}, \bibinfo
  {author} {\bibfnamefont {A.}~\bibnamefont {Bachtold}},\ and\ \bibinfo
  {author} {\bibfnamefont {P.}~\bibnamefont {Verlot}},\ }\bibfield  {title}
  {\bibinfo {title} {Optomechanics with a hybrid carbon nanotube resonator},\
  }\href {https://doi.org/10.1038/s41467-018-03097-z} {\bibfield  {journal}
  {\bibinfo  {journal} {Nat. Commun.}\ }\textbf {\bibinfo {volume} {9}},\
  \bibinfo {pages} {662} (\bibinfo {year} {2018})}\BibitemShut {NoStop}%
\bibitem [{\citenamefont {Fogliano}\ \emph
  {et~al.}(2021{\natexlab{a}})\citenamefont {Fogliano}, \citenamefont {Besga},
  \citenamefont {Reigue}, \citenamefont {Heringlake}, \citenamefont {Mercier~de
  L\'epinay}, \citenamefont {Vaneph}, \citenamefont {Reichel}, \citenamefont
  {Pigeau},\ and\ \citenamefont {Arcizet}}]{fogliano2021mapping}%
  \BibitemOpen
  \bibfield  {author} {\bibinfo {author} {\bibfnamefont {F.}~\bibnamefont
  {Fogliano}}, \bibinfo {author} {\bibfnamefont {B.}~\bibnamefont {Besga}},
  \bibinfo {author} {\bibfnamefont {A.}~\bibnamefont {Reigue}}, \bibinfo
  {author} {\bibfnamefont {P.}~\bibnamefont {Heringlake}}, \bibinfo {author}
  {\bibfnamefont {L.}~\bibnamefont {Mercier~de L\'epinay}}, \bibinfo {author}
  {\bibfnamefont {C.}~\bibnamefont {Vaneph}}, \bibinfo {author} {\bibfnamefont
  {J.}~\bibnamefont {Reichel}}, \bibinfo {author} {\bibfnamefont
  {B.}~\bibnamefont {Pigeau}},\ and\ \bibinfo {author} {\bibfnamefont
  {O.}~\bibnamefont {Arcizet}},\ }\bibfield  {title} {\bibinfo {title} {Mapping
  the cavity optomechanical interaction with subwavelength-sized ultrasensitive
  nanomechanical force sensors},\ }\href
  {https://doi.org/10.1103/PhysRevX.11.021009} {\bibfield  {journal} {\bibinfo
  {journal} {Phys. Rev. X}\ }\textbf {\bibinfo {volume} {11}},\ \bibinfo
  {pages} {021009} (\bibinfo {year} {2021}{\natexlab{a}})}\BibitemShut
  {NoStop}%
\bibitem [{\citenamefont {Ramos}\ \emph {et~al.}(2012)\citenamefont {Ramos},
  \citenamefont {Gil-Santos}, \citenamefont {Pini}, \citenamefont {Llorens},
  \citenamefont {Fern{\'a}ndez-Reg{\'u}lez}, \citenamefont {San~Paulo},
  \citenamefont {Calleja},\ and\ \citenamefont
  {Tamayo}}]{ramos2012optomechanics}%
  \BibitemOpen
  \bibfield  {author} {\bibinfo {author} {\bibfnamefont {D.}~\bibnamefont
  {Ramos}}, \bibinfo {author} {\bibfnamefont {E.}~\bibnamefont {Gil-Santos}},
  \bibinfo {author} {\bibfnamefont {V.}~\bibnamefont {Pini}}, \bibinfo {author}
  {\bibfnamefont {J.~M.}\ \bibnamefont {Llorens}}, \bibinfo {author}
  {\bibfnamefont {M.}~\bibnamefont {Fern{\'a}ndez-Reg{\'u}lez}}, \bibinfo
  {author} {\bibfnamefont {{\'A}.}~\bibnamefont {San~Paulo}}, \bibinfo {author}
  {\bibfnamefont {M.}~\bibnamefont {Calleja}},\ and\ \bibinfo {author}
  {\bibfnamefont {J.}~\bibnamefont {Tamayo}},\ }\bibfield  {title} {\bibinfo
  {title} {Optomechanics with silicon nanowires by harnessing confined
  electromagnetic modes},\ }\href {https://doi.org/10.1021/nl204002u}
  {\bibfield  {journal} {\bibinfo  {journal} {Nano Lett.}\ }\textbf {\bibinfo
  {volume} {12}},\ \bibinfo {pages} {932} (\bibinfo {year} {2012})}\BibitemShut
  {NoStop}%
\bibitem [{\citenamefont {Primo}\ \emph {et~al.}(2021)\citenamefont {Primo},
  \citenamefont {Kersul}, \citenamefont {Benevides}, \citenamefont {Carvalho},
  \citenamefont {M{\'e}nard}, \citenamefont {Frateschi}, \citenamefont
  {de~Assis}, \citenamefont {Wiederhecker},\ and\ \citenamefont
  {Mayer~Alegre}}]{primo2021accurate}%
  \BibitemOpen
  \bibfield  {author} {\bibinfo {author} {\bibfnamefont {A.~G.}\ \bibnamefont
  {Primo}}, \bibinfo {author} {\bibfnamefont {C.~M.}\ \bibnamefont {Kersul}},
  \bibinfo {author} {\bibfnamefont {R.}~\bibnamefont {Benevides}}, \bibinfo
  {author} {\bibfnamefont {N.~C.}\ \bibnamefont {Carvalho}}, \bibinfo {author}
  {\bibfnamefont {M.}~\bibnamefont {M{\'e}nard}}, \bibinfo {author}
  {\bibfnamefont {N.~C.}\ \bibnamefont {Frateschi}}, \bibinfo {author}
  {\bibfnamefont {P.-L.}\ \bibnamefont {de~Assis}}, \bibinfo {author}
  {\bibfnamefont {G.~S.}\ \bibnamefont {Wiederhecker}},\ and\ \bibinfo {author}
  {\bibfnamefont {T.~P.}\ \bibnamefont {Mayer~Alegre}},\ }\bibfield  {title}
  {\bibinfo {title} {{Accurate modeling and characterization of photothermal
  forces in optomechanics}},\ }\href {https://doi.org/10.1063/5.0055201}
  {\bibfield  {journal} {\bibinfo  {journal} {APL Photonics}\ }\textbf
  {\bibinfo {volume} {6}},\ \bibinfo {pages} {086101} (\bibinfo {year}
  {2021})}\BibitemShut {NoStop}%
\bibitem [{\citenamefont {Geitner}\ \emph {et~al.}(2017)\citenamefont
  {Geitner}, \citenamefont {Aguilar~Sandoval}, \citenamefont {Bertin},\ and\
  \citenamefont {Bellon}}]{Geitner2017}%
  \BibitemOpen
  \bibfield  {author} {\bibinfo {author} {\bibfnamefont {M.}~\bibnamefont
  {Geitner}}, \bibinfo {author} {\bibfnamefont {F.}~\bibnamefont
  {Aguilar~Sandoval}}, \bibinfo {author} {\bibfnamefont {E.}~\bibnamefont
  {Bertin}},\ and\ \bibinfo {author} {\bibfnamefont {L.}~\bibnamefont
  {Bellon}},\ }\bibfield  {title} {\bibinfo {title} {Low thermal fluctuations
  in a system heated out of equilibrium},\ }\href
  {https://doi.org/10.1103/PhysRevE.95.032138} {\bibfield  {journal} {\bibinfo
  {journal} {Phys. Rev. E}\ }\textbf {\bibinfo {volume} {95}},\ \bibinfo
  {pages} {032138} (\bibinfo {year} {2017})}\BibitemShut {NoStop}%
\bibitem [{\citenamefont {Shaniv}\ \emph {et~al.}(2023)\citenamefont {Shaniv},
  \citenamefont {Reetz},\ and\ \citenamefont {Regal}}]{shaniv2023direct}%
  \BibitemOpen
  \bibfield  {author} {\bibinfo {author} {\bibfnamefont {R.}~\bibnamefont
  {Shaniv}}, \bibinfo {author} {\bibfnamefont {C.}~\bibnamefont {Reetz}},\ and\
  \bibinfo {author} {\bibfnamefont {C.~A.}\ \bibnamefont {Regal}},\ }\bibfield
  {title} {\bibinfo {title} {Direct measurement of a spatially varying thermal
  bath using brownian motion},\ }\href
  {https://doi.org/10.1103/PhysRevResearch.5.043121} {\bibfield  {journal}
  {\bibinfo  {journal} {Phys. Rev. Res.}\ }\textbf {\bibinfo {volume} {5}},\
  \bibinfo {pages} {043121} (\bibinfo {year} {2023})}\BibitemShut {NoStop}%
\bibitem [{\citenamefont {Collin}\ \emph {et~al.}(2023)\citenamefont {Collin},
  \citenamefont {Golokolenov}, \citenamefont {Maillet}, \citenamefont
  {Saminadayar},\ and\ \citenamefont {Bourgeois}}]{Collin2023}%
  \BibitemOpen
  \bibfield  {author} {\bibinfo {author} {\bibfnamefont {E.}~\bibnamefont
  {Collin}}, \bibinfo {author} {\bibfnamefont {I.}~\bibnamefont {Golokolenov}},
  \bibinfo {author} {\bibfnamefont {O.}~\bibnamefont {Maillet}}, \bibinfo
  {author} {\bibfnamefont {L.}~\bibnamefont {Saminadayar}},\ and\ \bibinfo
  {author} {\bibfnamefont {O.}~\bibnamefont {Bourgeois}},\ }\bibfield  {title}
  {\bibinfo {title} {On the link between mechanics and thermal properties:
  mechanothermics},\ }\href {https://doi.org/10.1088/1367-2630/acc5a9}
  {\bibfield  {journal} {\bibinfo  {journal} {New J. Phys.}\ }\textbf {\bibinfo
  {volume} {25}},\ \bibinfo {pages} {043008} (\bibinfo {year}
  {2023})}\BibitemShut {NoStop}%
\bibitem [{\citenamefont {Lavrik}\ \emph {et~al.}(2004)\citenamefont {Lavrik},
  \citenamefont {Sepaniak},\ and\ \citenamefont {Datskos}}]{Lavrik2004}%
  \BibitemOpen
  \bibfield  {author} {\bibinfo {author} {\bibfnamefont {N.~V.}\ \bibnamefont
  {Lavrik}}, \bibinfo {author} {\bibfnamefont {M.~J.}\ \bibnamefont
  {Sepaniak}},\ and\ \bibinfo {author} {\bibfnamefont {P.~G.}\ \bibnamefont
  {Datskos}},\ }\bibfield  {title} {\bibinfo {title} {Cantilever transducers as
  a platform for chemical and biological sensors},\ }\href
  {https://doi.org/10.1063/1.1763252} {\bibfield  {journal} {\bibinfo
  {journal} {Rev. Sci. Instrum.}\ }\textbf {\bibinfo {volume} {75}},\ \bibinfo
  {pages} {2229} (\bibinfo {year} {2004})}\BibitemShut {NoStop}%
\bibitem [{\citenamefont {Garc\'{i}a}\ and\ \citenamefont
  {P\'{e}rez}(2002)}]{Garcia2002}%
  \BibitemOpen
  \bibfield  {author} {\bibinfo {author} {\bibfnamefont {R.}~\bibnamefont
  {Garc\'{i}a}}\ and\ \bibinfo {author} {\bibfnamefont {R.}~\bibnamefont
  {P\'{e}rez}},\ }\bibfield  {title} {\bibinfo {title} {Dynamic atomic force
  microscopy methods},\ }\href {https://doi.org/10.1016/S0167-5729(02)00077-8}
  {\bibfield  {journal} {\bibinfo  {journal} {Surf. Sci. Rep.}\ }\textbf
  {\bibinfo {volume} {47}},\ \bibinfo {pages} {197} (\bibinfo {year}
  {2002})}\BibitemShut {NoStop}%
\bibitem [{\citenamefont {Rossi}\ \emph {et~al.}(2017)\citenamefont {Rossi},
  \citenamefont {Braakman}, \citenamefont {Cadeddu}, \citenamefont {Vasyukov},
  \citenamefont {T{\"u}t{\"u}nc{\"u}oglu}, \citenamefont {Fontcuberta~i
  Morral},\ and\ \citenamefont {Poggio}}]{rossi2017vectorial}%
  \BibitemOpen
  \bibfield  {author} {\bibinfo {author} {\bibfnamefont {N.}~\bibnamefont
  {Rossi}}, \bibinfo {author} {\bibfnamefont {F.~R.}\ \bibnamefont {Braakman}},
  \bibinfo {author} {\bibfnamefont {D.}~\bibnamefont {Cadeddu}}, \bibinfo
  {author} {\bibfnamefont {D.}~\bibnamefont {Vasyukov}}, \bibinfo {author}
  {\bibfnamefont {G.}~\bibnamefont {T{\"u}t{\"u}nc{\"u}oglu}}, \bibinfo
  {author} {\bibfnamefont {A.}~\bibnamefont {Fontcuberta~i Morral}},\ and\
  \bibinfo {author} {\bibfnamefont {M.}~\bibnamefont {Poggio}},\ }\bibfield
  {title} {\bibinfo {title} {Vectorial scanning force microscopy using a
  nanowire sensor},\ }\href {https://doi.org/10.1038/nnano.2016.189} {\bibfield
   {journal} {\bibinfo  {journal} {Nat. Nanotech.}\ }\textbf {\bibinfo {volume}
  {12}},\ \bibinfo {pages} {150} (\bibinfo {year} {2017})}\BibitemShut
  {NoStop}%
\bibitem [{\citenamefont {de~L{\'e}pinay}\ \emph {et~al.}(2017)\citenamefont
  {de~L{\'e}pinay}, \citenamefont {Pigeau}, \citenamefont {Besga},
  \citenamefont {Vincent}, \citenamefont {Poncharal},\ and\ \citenamefont
  {Arcizet}}]{de2017universal}%
  \BibitemOpen
  \bibfield  {author} {\bibinfo {author} {\bibfnamefont {L.~M.}\ \bibnamefont
  {de~L{\'e}pinay}}, \bibinfo {author} {\bibfnamefont {B.}~\bibnamefont
  {Pigeau}}, \bibinfo {author} {\bibfnamefont {B.}~\bibnamefont {Besga}},
  \bibinfo {author} {\bibfnamefont {P.}~\bibnamefont {Vincent}}, \bibinfo
  {author} {\bibfnamefont {P.}~\bibnamefont {Poncharal}},\ and\ \bibinfo
  {author} {\bibfnamefont {O.}~\bibnamefont {Arcizet}},\ }\bibfield  {title}
  {\bibinfo {title} {A universal and ultrasensitive vectorial nanomechanical
  sensor for imaging 2d force fields},\ }\href
  {https://doi.org/10.1038/nnano.2016.193} {\bibfield  {journal} {\bibinfo
  {journal} {Nat. Nanotech.}\ }\textbf {\bibinfo {volume} {12}},\ \bibinfo
  {pages} {156} (\bibinfo {year} {2017})}\BibitemShut {NoStop}%
\bibitem [{\citenamefont {Buks}\ and\ \citenamefont {Roukes}(2000)}]{Buks2000}%
  \BibitemOpen
  \bibfield  {author} {\bibinfo {author} {\bibfnamefont {E.}~\bibnamefont
  {Buks}}\ and\ \bibinfo {author} {\bibfnamefont {M.~L.}\ \bibnamefont
  {Roukes}},\ }\bibfield  {title} {\bibinfo {title} {{Stiction, Adhesion Energy
  and the {C}asimir Effect in Micromechanical Systems}},\ }\href
  {https://doi.org/10.1103/PhysRevB.63.033402} {\bibfield  {journal} {\bibinfo
  {journal} {Phys. Rev. B}\ }\textbf {\bibinfo {volume} {63}},\ \bibinfo
  {pages} {033402} (\bibinfo {year} {2000})}\BibitemShut {NoStop}%
\bibitem [{\citenamefont {Pairis}\ \emph {et~al.}(2019)\citenamefont {Pairis},
  \citenamefont {Donatini}, \citenamefont {Hocevar}, \citenamefont {Tumanov},
  \citenamefont {Vaish}, \citenamefont {Claudon}, \citenamefont {Poizat},\ and\
  \citenamefont {Verlot}}]{pairis2019shot}%
  \BibitemOpen
  \bibfield  {author} {\bibinfo {author} {\bibfnamefont {S.}~\bibnamefont
  {Pairis}}, \bibinfo {author} {\bibfnamefont {F.}~\bibnamefont {Donatini}},
  \bibinfo {author} {\bibfnamefont {M.}~\bibnamefont {Hocevar}}, \bibinfo
  {author} {\bibfnamefont {D.}~\bibnamefont {Tumanov}}, \bibinfo {author}
  {\bibfnamefont {N.}~\bibnamefont {Vaish}}, \bibinfo {author} {\bibfnamefont
  {J.}~\bibnamefont {Claudon}}, \bibinfo {author} {\bibfnamefont {J.-P.}\
  \bibnamefont {Poizat}},\ and\ \bibinfo {author} {\bibfnamefont
  {P.}~\bibnamefont {Verlot}},\ }\bibfield  {title} {\bibinfo {title}
  {Shot-noise-limited nanomechanical detection and radiation pressure
  backaction from an electron beam},\ }\href
  {https://doi.org/10.1103/PhysRevLett.122.083603} {\bibfield  {journal}
  {\bibinfo  {journal} {Phys. Rev. Lett.}\ }\textbf {\bibinfo {volume} {122}},\
  \bibinfo {pages} {083603} (\bibinfo {year} {2019})}\BibitemShut {NoStop}%
\bibitem [{\citenamefont {Cretu}\ \emph {et~al.}(2022)\citenamefont {Cretu},
  \citenamefont {Zhang},\ and\ \citenamefont {Kimoto}}]{cretu2022direct}%
  \BibitemOpen
  \bibfield  {author} {\bibinfo {author} {\bibfnamefont {O.}~\bibnamefont
  {Cretu}}, \bibinfo {author} {\bibfnamefont {H.}~\bibnamefont {Zhang}},\ and\
  \bibinfo {author} {\bibfnamefont {K.}~\bibnamefont {Kimoto}},\ }\bibfield
  {title} {\bibinfo {title} {Direct observation of thermal vibration modes
  using frequency-selective electron microscopy},\ }\href
  {https://doi.org/10.1021/acs.nanolett.2c03762} {\bibfield  {journal}
  {\bibinfo  {journal} {Nano Lett.}\ }\textbf {\bibinfo {volume} {22}},\
  \bibinfo {pages} {10034} (\bibinfo {year} {2022})}\BibitemShut {NoStop}%
\bibitem [{\citenamefont {Sanii}\ and\ \citenamefont
  {Ashby}(2010)}]{sanii2010high}%
  \BibitemOpen
  \bibfield  {author} {\bibinfo {author} {\bibfnamefont {B.}~\bibnamefont
  {Sanii}}\ and\ \bibinfo {author} {\bibfnamefont {P.~D.}\ \bibnamefont
  {Ashby}},\ }\bibfield  {title} {\bibinfo {title} {High sensitivity deflection
  detection of nanowires},\ }\href
  {https://doi.org/10.1103/PhysRevLett.104.147203} {\bibfield  {journal}
  {\bibinfo  {journal} {Phys. Rev. Lett.}\ }\textbf {\bibinfo {volume} {104}},\
  \bibinfo {pages} {147203} (\bibinfo {year} {2010})}\BibitemShut {NoStop}%
\bibitem [{\citenamefont {Paolino}\ \emph {et~al.}(2013)\citenamefont
  {Paolino}, \citenamefont {Aguilar~Sandoval},\ and\ \citenamefont
  {Bellon}}]{paolino2013quadrature}%
  \BibitemOpen
  \bibfield  {author} {\bibinfo {author} {\bibfnamefont {P.}~\bibnamefont
  {Paolino}}, \bibinfo {author} {\bibfnamefont {F.~A.}\ \bibnamefont
  {Aguilar~Sandoval}},\ and\ \bibinfo {author} {\bibfnamefont {L.}~\bibnamefont
  {Bellon}},\ }\bibfield  {title} {\bibinfo {title} {Quadrature phase
  interferometer for high resolution force spectroscopy},\ }\href
  {https://doi.org/10.1063/1.4819743} {\bibfield  {journal} {\bibinfo
  {journal} {Rev. Sci. Instrum.}\ }\textbf {\bibinfo {volume} {84}},\ \bibinfo
  {pages} {095001} (\bibinfo {year} {2013})}\BibitemShut {NoStop}%
\bibitem [{\citenamefont {De~Liberato}\ \emph {et~al.}(2011)\citenamefont
  {De~Liberato}, \citenamefont {Lambert},\ and\ \citenamefont
  {Nori}}]{DeLiberato2011}%
  \BibitemOpen
  \bibfield  {author} {\bibinfo {author} {\bibfnamefont {S.}~\bibnamefont
  {De~Liberato}}, \bibinfo {author} {\bibfnamefont {N.}~\bibnamefont
  {Lambert}},\ and\ \bibinfo {author} {\bibfnamefont {F.}~\bibnamefont
  {Nori}},\ }\bibfield  {title} {\bibinfo {title} {Quantum noise in
  photothermal cooling},\ }\href {https://doi.org/10.1103/PhysRevA.83.033809}
  {\bibfield  {journal} {\bibinfo  {journal} {Phys. Rev. A}\ }\textbf {\bibinfo
  {volume} {83}},\ \bibinfo {pages} {033809} (\bibinfo {year}
  {2011})}\BibitemShut {NoStop}%
\bibitem [{\citenamefont {Tsioutsios}\ \emph {et~al.}(2017)\citenamefont
  {Tsioutsios}, \citenamefont {Tavernarakis}, \citenamefont {Osmond},
  \citenamefont {Verlot},\ and\ \citenamefont {Bachtold}}]{tsioutsios2017real}%
  \BibitemOpen
  \bibfield  {author} {\bibinfo {author} {\bibfnamefont {I.}~\bibnamefont
  {Tsioutsios}}, \bibinfo {author} {\bibfnamefont {A.}~\bibnamefont
  {Tavernarakis}}, \bibinfo {author} {\bibfnamefont {J.}~\bibnamefont
  {Osmond}}, \bibinfo {author} {\bibfnamefont {P.}~\bibnamefont {Verlot}},\
  and\ \bibinfo {author} {\bibfnamefont {A.}~\bibnamefont {Bachtold}},\
  }\bibfield  {title} {\bibinfo {title} {Real-time measurement of nanotube
  resonator fluctuations in an electron microscope},\ }\href
  {https://doi.org/10.1021/acs.nanolett.9b02351} {\bibfield  {journal}
  {\bibinfo  {journal} {Nano Lett.}\ }\textbf {\bibinfo {volume} {17}},\
  \bibinfo {pages} {1748} (\bibinfo {year} {2017})}\BibitemShut {NoStop}%
\bibitem [{\citenamefont {Allegrini}\ \emph {et~al.}(1992)\citenamefont
  {Allegrini}, \citenamefont {Ascoli}, \citenamefont {Baschieri}, \citenamefont
  {Dinelli}, \citenamefont {Frediani}, \citenamefont {Lio},\ and\ \citenamefont
  {Mariani}}]{Allegrini-1992}%
  \BibitemOpen
  \bibfield  {author} {\bibinfo {author} {\bibfnamefont {M.}~\bibnamefont
  {Allegrini}}, \bibinfo {author} {\bibfnamefont {C.}~\bibnamefont {Ascoli}},
  \bibinfo {author} {\bibfnamefont {P.}~\bibnamefont {Baschieri}}, \bibinfo
  {author} {\bibfnamefont {F.}~\bibnamefont {Dinelli}}, \bibinfo {author}
  {\bibfnamefont {C.}~\bibnamefont {Frediani}}, \bibinfo {author}
  {\bibfnamefont {A.}~\bibnamefont {Lio}},\ and\ \bibinfo {author}
  {\bibfnamefont {T.}~\bibnamefont {Mariani}},\ }\bibfield  {title} {\bibinfo
  {title} {Laser thermal effects on atomic force microscope cantilevers},\
  }\href {https://doi.org/10.1016/0304-3991(92)90295-U} {\bibfield  {journal}
  {\bibinfo  {journal} {Ultramicroscopy}\ }\textbf {\bibinfo {volume}
  {42--44}},\ \bibinfo {pages} {371} (\bibinfo {year} {1992})}\BibitemShut
  {NoStop}%
\bibitem [{Note1()}]{Note1}%
  \BibitemOpen
  \bibinfo {note} {Just as the static temperature field depends linearly on the
  mean power $\protect \overline {P}$, the static deflection profile $\protect
  \bar \xi (y)$ can change due to the static temperature field $\protect
  \overline {T}(y)$. This effect will just offset the average position of the
  oscillator, and is taken into account by defining $\xi $ as the dynamic
  deviation to $\protect \bar \xi $. When tuning the lateral position $x_p$ of
  the probe to maximize sensitivity, one should recover the same level of
  absorption $a$ for any imposed $\protect \overline {P}$. In the linear
  framework that we use, this static deflection does not have any effect on the
  dynamics.}\BibitemShut {Stop}%
\bibitem [{\citenamefont {Barton}\ \emph {et~al.}(2012)\citenamefont {Barton},
  \citenamefont {Storch}, \citenamefont {Adiga}, \citenamefont {Sakakibara},
  \citenamefont {Cipriany}, \citenamefont {Ilic}, \citenamefont {Wang},
  \citenamefont {Ong}, \citenamefont {McEuen}, \citenamefont {Parpia},\ and\
  \citenamefont {Craighead}}]{barton2012}%
  \BibitemOpen
  \bibfield  {author} {\bibinfo {author} {\bibfnamefont {R.~A.}\ \bibnamefont
  {Barton}}, \bibinfo {author} {\bibfnamefont {I.~R.}\ \bibnamefont {Storch}},
  \bibinfo {author} {\bibfnamefont {V.~P.}\ \bibnamefont {Adiga}}, \bibinfo
  {author} {\bibfnamefont {R.}~\bibnamefont {Sakakibara}}, \bibinfo {author}
  {\bibfnamefont {B.~R.}\ \bibnamefont {Cipriany}}, \bibinfo {author}
  {\bibfnamefont {B.}~\bibnamefont {Ilic}}, \bibinfo {author} {\bibfnamefont
  {S.~P.}\ \bibnamefont {Wang}}, \bibinfo {author} {\bibfnamefont
  {P.}~\bibnamefont {Ong}}, \bibinfo {author} {\bibfnamefont {P.~L.}\
  \bibnamefont {McEuen}}, \bibinfo {author} {\bibfnamefont {J.~M.}\
  \bibnamefont {Parpia}},\ and\ \bibinfo {author} {\bibfnamefont {H.~G.}\
  \bibnamefont {Craighead}},\ }\bibfield  {title} {\bibinfo {title}
  {Photothermal self-oscillation and laser cooling of graphene optomechanical
  systems},\ }\href {https://doi.org/10.1021/nl302036x} {\bibfield  {journal}
  {\bibinfo  {journal} {Nano Lett.}\ }\textbf {\bibinfo {volume} {12}},\
  \bibinfo {pages} {4681} (\bibinfo {year} {2012})}\BibitemShut {NoStop}%
\bibitem [{\citenamefont {Ashkin}\ \emph {et~al.}(1986)\citenamefont {Ashkin},
  \citenamefont {Dziedzic}, \citenamefont {Bjorkholm},\ and\ \citenamefont
  {Chu}}]{ashkin1986observation}%
  \BibitemOpen
  \bibfield  {author} {\bibinfo {author} {\bibfnamefont {A.}~\bibnamefont
  {Ashkin}}, \bibinfo {author} {\bibfnamefont {J.~M.}\ \bibnamefont
  {Dziedzic}}, \bibinfo {author} {\bibfnamefont {J.}~\bibnamefont
  {Bjorkholm}},\ and\ \bibinfo {author} {\bibfnamefont {S.}~\bibnamefont
  {Chu}},\ }\bibfield  {title} {\bibinfo {title} {Observation of a single-beam
  gradient force optical trap for dielectric particles},\ }\href
  {https://doi.org/10.1364/OL.11.000288} {\bibfield  {journal} {\bibinfo
  {journal} {Opt. Lett.}\ }\textbf {\bibinfo {volume} {11}},\ \bibinfo {pages}
  {288} (\bibinfo {year} {1986})}\BibitemShut {NoStop}%
\bibitem [{\citenamefont {Jain}\ \emph {et~al.}(2016)\citenamefont {Jain},
  \citenamefont {Gieseler}, \citenamefont {Moritz}, \citenamefont {Dellago},
  \citenamefont {Quidant},\ and\ \citenamefont {Novotny}}]{jain2016direct}%
  \BibitemOpen
  \bibfield  {author} {\bibinfo {author} {\bibfnamefont {V.}~\bibnamefont
  {Jain}}, \bibinfo {author} {\bibfnamefont {J.}~\bibnamefont {Gieseler}},
  \bibinfo {author} {\bibfnamefont {C.}~\bibnamefont {Moritz}}, \bibinfo
  {author} {\bibfnamefont {C.}~\bibnamefont {Dellago}}, \bibinfo {author}
  {\bibfnamefont {R.}~\bibnamefont {Quidant}},\ and\ \bibinfo {author}
  {\bibfnamefont {L.}~\bibnamefont {Novotny}},\ }\bibfield  {title} {\bibinfo
  {title} {Direct measurement of photon recoil from a levitated nanoparticle},\
  }\href {https://doi.org/10.1103/PhysRevLett.116.243601} {\bibfield  {journal}
  {\bibinfo  {journal} {Phys. Rev. Lett.}\ }\textbf {\bibinfo {volume} {116}},\
  \bibinfo {pages} {243601} (\bibinfo {year} {2016})}\BibitemShut {NoStop}%
\bibitem [{\citenamefont {Anetsberger}\ \emph {et~al.}(2009)\citenamefont
  {Anetsberger}, \citenamefont {Arcizet}, \citenamefont {Unterreithmeier},
  \citenamefont {Rivi{\`e}re}, \citenamefont {Schliesser}, \citenamefont
  {Weig}, \citenamefont {Kotthaus},\ and\ \citenamefont
  {Kippenberg}}]{anetsberger2009near}%
  \BibitemOpen
  \bibfield  {author} {\bibinfo {author} {\bibfnamefont {G.}~\bibnamefont
  {Anetsberger}}, \bibinfo {author} {\bibfnamefont {O.}~\bibnamefont
  {Arcizet}}, \bibinfo {author} {\bibfnamefont {Q.~P.}\ \bibnamefont
  {Unterreithmeier}}, \bibinfo {author} {\bibfnamefont {R.}~\bibnamefont
  {Rivi{\`e}re}}, \bibinfo {author} {\bibfnamefont {A.}~\bibnamefont
  {Schliesser}}, \bibinfo {author} {\bibfnamefont {E.~M.}\ \bibnamefont
  {Weig}}, \bibinfo {author} {\bibfnamefont {J.~P.}\ \bibnamefont {Kotthaus}},\
  and\ \bibinfo {author} {\bibfnamefont {T.~J.}\ \bibnamefont {Kippenberg}},\
  }\bibfield  {title} {\bibinfo {title} {Near-field cavity optomechanics with
  nanomechanical oscillators},\ }\href {https://doi.org/10.1038/nphys1425}
  {\bibfield  {journal} {\bibinfo  {journal} {Nat. Phys.}\ }\textbf {\bibinfo
  {volume} {5}},\ \bibinfo {pages} {909} (\bibinfo {year} {2009})}\BibitemShut
  {NoStop}%
\bibitem [{\citenamefont {Arcizet}\ \emph {et~al.}(2011)\citenamefont
  {Arcizet}, \citenamefont {Jacques}, \citenamefont {Siria}, \citenamefont
  {Poncharal}, \citenamefont {Vincent},\ and\ \citenamefont
  {Seidelin}}]{arcizet2011single}%
  \BibitemOpen
  \bibfield  {author} {\bibinfo {author} {\bibfnamefont {O.}~\bibnamefont
  {Arcizet}}, \bibinfo {author} {\bibfnamefont {V.}~\bibnamefont {Jacques}},
  \bibinfo {author} {\bibfnamefont {A.}~\bibnamefont {Siria}}, \bibinfo
  {author} {\bibfnamefont {P.}~\bibnamefont {Poncharal}}, \bibinfo {author}
  {\bibfnamefont {P.}~\bibnamefont {Vincent}},\ and\ \bibinfo {author}
  {\bibfnamefont {S.}~\bibnamefont {Seidelin}},\ }\bibfield  {title} {\bibinfo
  {title} {A single nitrogen-vacancy defect coupled to a nanomechanical
  oscillator},\ }\href {https://doi.org/10.1038/nphys2070} {\bibfield
  {journal} {\bibinfo  {journal} {Nat. Phys.}\ }\textbf {\bibinfo {volume}
  {7}},\ \bibinfo {pages} {879} (\bibinfo {year} {2011})}\BibitemShut {NoStop}%
\bibitem [{\citenamefont {Teissier}\ \emph {et~al.}(2014)\citenamefont
  {Teissier}, \citenamefont {Barfuss}, \citenamefont {Appel}, \citenamefont
  {Neu},\ and\ \citenamefont {Maletinsky}}]{teissier2014strain}%
  \BibitemOpen
  \bibfield  {author} {\bibinfo {author} {\bibfnamefont {J.}~\bibnamefont
  {Teissier}}, \bibinfo {author} {\bibfnamefont {A.}~\bibnamefont {Barfuss}},
  \bibinfo {author} {\bibfnamefont {P.}~\bibnamefont {Appel}}, \bibinfo
  {author} {\bibfnamefont {E.}~\bibnamefont {Neu}},\ and\ \bibinfo {author}
  {\bibfnamefont {P.}~\bibnamefont {Maletinsky}},\ }\bibfield  {title}
  {\bibinfo {title} {Strain coupling of a nitrogen-vacancy center spin to a
  diamond mechanical oscillator},\ }\href
  {https://doi.org/10.1103/PhysRevLett.113.020503} {\bibfield  {journal}
  {\bibinfo  {journal} {Phys. Rev. Lett.}\ }\textbf {\bibinfo {volume} {113}},\
  \bibinfo {pages} {020503} (\bibinfo {year} {2014})}\BibitemShut {NoStop}%
\bibitem [{\citenamefont {Kettler}\ \emph {et~al.}(2021)\citenamefont
  {Kettler}, \citenamefont {Vaish}, \citenamefont {de~L{\'e}pinay},
  \citenamefont {Besga}, \citenamefont {de~Assis}, \citenamefont {Bourgeois},
  \citenamefont {Auff{\`e}ves}, \citenamefont {Richard}, \citenamefont
  {Claudon}, \citenamefont {G{\'e}rard}, \citenamefont {Pigeau}, \citenamefont
  {Arcizet}, \citenamefont {Verlot},\ and\ \citenamefont
  {Poizat}}]{kettler2021inducing}%
  \BibitemOpen
  \bibfield  {author} {\bibinfo {author} {\bibfnamefont {J.}~\bibnamefont
  {Kettler}}, \bibinfo {author} {\bibfnamefont {N.}~\bibnamefont {Vaish}},
  \bibinfo {author} {\bibfnamefont {L.~M.}\ \bibnamefont {de~L{\'e}pinay}},
  \bibinfo {author} {\bibfnamefont {B.}~\bibnamefont {Besga}}, \bibinfo
  {author} {\bibfnamefont {P.-L.}\ \bibnamefont {de~Assis}}, \bibinfo {author}
  {\bibfnamefont {O.}~\bibnamefont {Bourgeois}}, \bibinfo {author}
  {\bibfnamefont {A.}~\bibnamefont {Auff{\`e}ves}}, \bibinfo {author}
  {\bibfnamefont {M.}~\bibnamefont {Richard}}, \bibinfo {author} {\bibfnamefont
  {J.}~\bibnamefont {Claudon}}, \bibinfo {author} {\bibfnamefont {J.-M.}\
  \bibnamefont {G{\'e}rard}}, \bibinfo {author} {\bibfnamefont
  {B.}~\bibnamefont {Pigeau}}, \bibinfo {author} {\bibfnamefont
  {O.}~\bibnamefont {Arcizet}}, \bibinfo {author} {\bibfnamefont
  {P.}~\bibnamefont {Verlot}},\ and\ \bibinfo {author} {\bibfnamefont {J.-P.}\
  \bibnamefont {Poizat}},\ }\bibfield  {title} {\bibinfo {title} {Inducing
  micromechanical motion by optical excitation of a single quantum dot},\
  }\href {https://doi.org/10.1038/s41565-020-00814-y} {\bibfield  {journal}
  {\bibinfo  {journal} {Nat. Nanotech.}\ }\textbf {\bibinfo {volume} {16}},\
  \bibinfo {pages} {283} (\bibinfo {year} {2021})}\BibitemShut {NoStop}%
\bibitem [{\citenamefont {Komori}\ \emph {et~al.}(2018)\citenamefont {Komori},
  \citenamefont {Enomoto}, \citenamefont {Takeda}, \citenamefont {Michimura},
  \citenamefont {Somiya}, \citenamefont {Ando},\ and\ \citenamefont
  {Ballmer}}]{Komori2018}%
  \BibitemOpen
  \bibfield  {author} {\bibinfo {author} {\bibfnamefont {K.}~\bibnamefont
  {Komori}}, \bibinfo {author} {\bibfnamefont {Y.}~\bibnamefont {Enomoto}},
  \bibinfo {author} {\bibfnamefont {H.}~\bibnamefont {Takeda}}, \bibinfo
  {author} {\bibfnamefont {Y.}~\bibnamefont {Michimura}}, \bibinfo {author}
  {\bibfnamefont {K.}~\bibnamefont {Somiya}}, \bibinfo {author} {\bibfnamefont
  {M.}~\bibnamefont {Ando}},\ and\ \bibinfo {author} {\bibfnamefont {S.~W.}\
  \bibnamefont {Ballmer}},\ }\bibfield  {title} {\bibinfo {title} {Direct
  approach for the fluctuation-dissipation theorem under nonequilibrium
  steady-state conditions},\ }\href
  {https://doi.org/10.1103/PhysRevD.97.102001} {\bibfield  {journal} {\bibinfo
  {journal} {Phys. Rev. D}\ }\textbf {\bibinfo {volume} {{97}}},\ \bibinfo
  {pages} {102001} (\bibinfo {year} {2018})}\BibitemShut {NoStop}%
\bibitem [{\citenamefont {Fontana}\ \emph {et~al.}(2020)\citenamefont
  {Fontana}, \citenamefont {Pedurand},\ and\ \citenamefont
  {Bellon}}]{Fontana2020}%
  \BibitemOpen
  \bibfield  {author} {\bibinfo {author} {\bibfnamefont {A.}~\bibnamefont
  {Fontana}}, \bibinfo {author} {\bibfnamefont {R.}~\bibnamefont {Pedurand}},\
  and\ \bibinfo {author} {\bibfnamefont {L.}~\bibnamefont {Bellon}},\
  }\bibfield  {title} {\bibinfo {title} {Extended equipartition in a mechanical
  system subject to a heat flow: the case of localised dissipation},\ }\href
  {https://doi.org/10.1088/1742-5468/ab97b1} {\bibfield  {journal} {\bibinfo
  {journal} {J. Stat. Mech.}\ }\textbf {\bibinfo {volume} {2020}},\ \bibinfo
  {pages} {073206} (\bibinfo {year} {2020})}\BibitemShut {NoStop}%
\bibitem [{\citenamefont {Fontana}\ and\ \citenamefont
  {Bellon}(2023)}]{Fontana2023}%
  \BibitemOpen
  \bibfield  {author} {\bibinfo {author} {\bibfnamefont {A.}~\bibnamefont
  {Fontana}}\ and\ \bibinfo {author} {\bibfnamefont {L.}~\bibnamefont
  {Bellon}},\ }\bibfield  {title} {\bibinfo {title} {Linking fluctuation and
  dissipation in spatially extended out-of-equilibrium systems},\ }\href
  {https://doi.org/10.1103/PhysRevE.107.034118} {\bibfield  {journal} {\bibinfo
   {journal} {Phys. Rev. E}\ }\textbf {\bibinfo {volume} {107}},\ \bibinfo
  {pages} {034118} (\bibinfo {year} {2023})}\BibitemShut {NoStop}%
\bibitem [{\citenamefont {Pigeau}\ \emph {et~al.}(2015)\citenamefont {Pigeau},
  \citenamefont {Rohr}, \citenamefont {Mercier~de L{\'e}pinay}, \citenamefont
  {Gloppe}, \citenamefont {Jacques},\ and\ \citenamefont
  {Arcizet}}]{pigeau2015observation}%
  \BibitemOpen
  \bibfield  {author} {\bibinfo {author} {\bibfnamefont {B.}~\bibnamefont
  {Pigeau}}, \bibinfo {author} {\bibfnamefont {S.}~\bibnamefont {Rohr}},
  \bibinfo {author} {\bibfnamefont {L.}~\bibnamefont {Mercier~de L{\'e}pinay}},
  \bibinfo {author} {\bibfnamefont {A.}~\bibnamefont {Gloppe}}, \bibinfo
  {author} {\bibfnamefont {V.}~\bibnamefont {Jacques}},\ and\ \bibinfo {author}
  {\bibfnamefont {O.}~\bibnamefont {Arcizet}},\ }\bibfield  {title} {\bibinfo
  {title} {Observation of a phononic mollow triplet in a multimode hybrid
  spin-nanomechanical system},\ }\href {https://doi.org/10.1038/ncomms9603}
  {\bibfield  {journal} {\bibinfo  {journal} {Nat. Commun.}\ }\textbf {\bibinfo
  {volume} {6}},\ \bibinfo {pages} {8603} (\bibinfo {year} {2015})}\BibitemShut
  {NoStop}%
\bibitem [{\citenamefont {Budakian}\ \emph {et~al.}(2024)\citenamefont
  {Budakian}, \citenamefont {Finkler}, \citenamefont {Eichler}, \citenamefont
  {Poggio}, \citenamefont {Degen}, \citenamefont {Tabatabaei}, \citenamefont
  {Lee}, \citenamefont {Hammel}, \citenamefont {Eugene}, \citenamefont
  {Taminiau}, \citenamefont {Walsworth}, \citenamefont {London}, \citenamefont
  {Jayich}, \citenamefont {Ajoy}, \citenamefont {Pillai}, \citenamefont
  {Wrachtrup}, \citenamefont {Jelezko}, \citenamefont {Bae}, \citenamefont
  {Heinrich}, \citenamefont {Ast}, \citenamefont {Bertet}, \citenamefont
  {Cappellaro}, \citenamefont {Bonato}, \citenamefont {Altmann},\ and\
  \citenamefont {Gauger}}]{budakian2024roadmap}%
  \BibitemOpen
  \bibfield  {author} {\bibinfo {author} {\bibfnamefont {R.}~\bibnamefont
  {Budakian}}, \bibinfo {author} {\bibfnamefont {A.}~\bibnamefont {Finkler}},
  \bibinfo {author} {\bibfnamefont {A.}~\bibnamefont {Eichler}}, \bibinfo
  {author} {\bibfnamefont {M.}~\bibnamefont {Poggio}}, \bibinfo {author}
  {\bibfnamefont {C.~L.}\ \bibnamefont {Degen}}, \bibinfo {author}
  {\bibfnamefont {S.}~\bibnamefont {Tabatabaei}}, \bibinfo {author}
  {\bibfnamefont {I.}~\bibnamefont {Lee}}, \bibinfo {author} {\bibfnamefont
  {P.~C.}\ \bibnamefont {Hammel}}, \bibinfo {author} {\bibfnamefont {S.~P.}\
  \bibnamefont {Eugene}}, \bibinfo {author} {\bibfnamefont {T.~H.}\
  \bibnamefont {Taminiau}}, \bibinfo {author} {\bibfnamefont {R.~L.}\
  \bibnamefont {Walsworth}}, \bibinfo {author} {\bibfnamefont {P.}~\bibnamefont
  {London}}, \bibinfo {author} {\bibfnamefont {A.~B.}\ \bibnamefont {Jayich}},
  \bibinfo {author} {\bibfnamefont {A.}~\bibnamefont {Ajoy}}, \bibinfo {author}
  {\bibfnamefont {A.}~\bibnamefont {Pillai}}, \bibinfo {author} {\bibfnamefont
  {J.}~\bibnamefont {Wrachtrup}}, \bibinfo {author} {\bibfnamefont
  {F.}~\bibnamefont {Jelezko}}, \bibinfo {author} {\bibfnamefont
  {Y.}~\bibnamefont {Bae}}, \bibinfo {author} {\bibfnamefont {A.~J.}\
  \bibnamefont {Heinrich}}, \bibinfo {author} {\bibfnamefont {C.~R.}\
  \bibnamefont {Ast}}, \bibinfo {author} {\bibfnamefont {P.}~\bibnamefont
  {Bertet}}, \bibinfo {author} {\bibfnamefont {P.}~\bibnamefont {Cappellaro}},
  \bibinfo {author} {\bibfnamefont {C.}~\bibnamefont {Bonato}}, \bibinfo
  {author} {\bibfnamefont {Y.}~\bibnamefont {Altmann}},\ and\ \bibinfo {author}
  {\bibfnamefont {E.}~\bibnamefont {Gauger}},\ }\bibfield  {title} {\bibinfo
  {title} {Roadmap on nanoscale magnetic resonance imaging},\ }\href
  {https://doi.org/10.1088/1361-6528/ad4b23} {\bibfield  {journal} {\bibinfo
  {journal} {Nanotechnology}\ }\textbf {\bibinfo {volume} {35}},\ \bibinfo
  {pages} {412001} (\bibinfo {year} {2024})}\BibitemShut {NoStop}%
\bibitem [{\citenamefont {Sahafi}\ \emph {et~al.}(2020)\citenamefont {Sahafi},
  \citenamefont {Rose}, \citenamefont {Jordan}, \citenamefont {Yager},
  \citenamefont {Piscitelli},\ and\ \citenamefont
  {Budakian}}]{sahafi2019ultralow}%
  \BibitemOpen
  \bibfield  {author} {\bibinfo {author} {\bibfnamefont {P.}~\bibnamefont
  {Sahafi}}, \bibinfo {author} {\bibfnamefont {W.}~\bibnamefont {Rose}},
  \bibinfo {author} {\bibfnamefont {A.}~\bibnamefont {Jordan}}, \bibinfo
  {author} {\bibfnamefont {B.}~\bibnamefont {Yager}}, \bibinfo {author}
  {\bibfnamefont {M.}~\bibnamefont {Piscitelli}},\ and\ \bibinfo {author}
  {\bibfnamefont {R.}~\bibnamefont {Budakian}},\ }\bibfield  {title} {\bibinfo
  {title} {Ultralow dissipation patterned silicon nanowire arrays for scanning
  probe microscopy},\ }\href {https://doi.org/10.1021/acs.nanolett.9b03668}
  {\bibfield  {journal} {\bibinfo  {journal} {Nano Lett.}\ }\textbf {\bibinfo
  {volume} {20}},\ \bibinfo {pages} {218} (\bibinfo {year} {2020})}\BibitemShut
  {NoStop}%
\bibitem [{\citenamefont {Fogliano}\ \emph
  {et~al.}(2021{\natexlab{b}})\citenamefont {Fogliano}, \citenamefont {Besga},
  \citenamefont {Reigue}, \citenamefont {Mercier~de L{\'e}pinay}, \citenamefont
  {Heringlake}, \citenamefont {Gouriou}, \citenamefont {Eyraud}, \citenamefont
  {Wernsdorfer}, \citenamefont {Pigeau},\ and\ \citenamefont
  {Arcizet}}]{fogliano2021ultrasensitive}%
  \BibitemOpen
  \bibfield  {author} {\bibinfo {author} {\bibfnamefont {F.}~\bibnamefont
  {Fogliano}}, \bibinfo {author} {\bibfnamefont {B.}~\bibnamefont {Besga}},
  \bibinfo {author} {\bibfnamefont {A.}~\bibnamefont {Reigue}}, \bibinfo
  {author} {\bibfnamefont {L.}~\bibnamefont {Mercier~de L{\'e}pinay}}, \bibinfo
  {author} {\bibfnamefont {P.}~\bibnamefont {Heringlake}}, \bibinfo {author}
  {\bibfnamefont {C.}~\bibnamefont {Gouriou}}, \bibinfo {author} {\bibfnamefont
  {E.}~\bibnamefont {Eyraud}}, \bibinfo {author} {\bibfnamefont
  {W.}~\bibnamefont {Wernsdorfer}}, \bibinfo {author} {\bibfnamefont
  {B.}~\bibnamefont {Pigeau}},\ and\ \bibinfo {author} {\bibfnamefont
  {O.}~\bibnamefont {Arcizet}},\ }\bibfield  {title} {\bibinfo {title}
  {Ultrasensitive nano-optomechanical force sensor operated at dilution
  temperatures},\ }\href {https://doi.org/10.1038/s41467-021-24318-y}
  {\bibfield  {journal} {\bibinfo  {journal} {Nat. Commun.}\ }\textbf {\bibinfo
  {volume} {12}},\ \bibinfo {pages} {4124} (\bibinfo {year}
  {2021}{\natexlab{b}})}\BibitemShut {NoStop}%
\bibitem [{\citenamefont {Valent{\'\i}n}\ \emph {et~al.}(2013)\citenamefont
  {Valent{\'\i}n}, \citenamefont {Betancourt}, \citenamefont {Fonseca},
  \citenamefont {Pettes}, \citenamefont {Shi}, \citenamefont {Soszy{\'n}ski},\
  and\ \citenamefont {Huczko}}]{valentin2013comprehensive}%
  \BibitemOpen
  \bibfield  {author} {\bibinfo {author} {\bibfnamefont {L.~A.}\ \bibnamefont
  {Valent{\'\i}n}}, \bibinfo {author} {\bibfnamefont {J.}~\bibnamefont
  {Betancourt}}, \bibinfo {author} {\bibfnamefont {L.~F.}\ \bibnamefont
  {Fonseca}}, \bibinfo {author} {\bibfnamefont {M.~T.}\ \bibnamefont {Pettes}},
  \bibinfo {author} {\bibfnamefont {L.}~\bibnamefont {Shi}}, \bibinfo {author}
  {\bibfnamefont {M.}~\bibnamefont {Soszy{\'n}ski}},\ and\ \bibinfo {author}
  {\bibfnamefont {A.}~\bibnamefont {Huczko}},\ }\bibfield  {title} {\bibinfo
  {title} {{A comprehensive study of thermoelectric and transport properties
  of$\beta$-silicon carbide nanowires}},\ }\href
  {https://doi.org/10.1063/1.4829924} {\bibfield  {journal} {\bibinfo
  {journal} {J. Appl. Phys.}\ }\textbf {\bibinfo {volume} {114}},\ \bibinfo
  {pages} {184301} (\bibinfo {year} {2013})}\BibitemShut {NoStop}%
\bibitem [{\citenamefont {Ghadimi}\ \emph {et~al.}(2018)\citenamefont
  {Ghadimi}, \citenamefont {Fedorov}, \citenamefont {Engelsen}, \citenamefont
  {Bereyhi}, \citenamefont {Schilling}, \citenamefont {Wilson},\ and\
  \citenamefont {Kippenberg}}]{ghadimi2018elastic}%
  \BibitemOpen
  \bibfield  {author} {\bibinfo {author} {\bibfnamefont {A.~H.}\ \bibnamefont
  {Ghadimi}}, \bibinfo {author} {\bibfnamefont {S.~A.}\ \bibnamefont
  {Fedorov}}, \bibinfo {author} {\bibfnamefont {N.~J.}\ \bibnamefont
  {Engelsen}}, \bibinfo {author} {\bibfnamefont {M.~J.}\ \bibnamefont
  {Bereyhi}}, \bibinfo {author} {\bibfnamefont {R.}~\bibnamefont {Schilling}},
  \bibinfo {author} {\bibfnamefont {D.~J.}\ \bibnamefont {Wilson}},\ and\
  \bibinfo {author} {\bibfnamefont {T.~J.}\ \bibnamefont {Kippenberg}},\
  }\bibfield  {title} {\bibinfo {title} {Elastic strain engineering for
  ultralow mechanical dissipation},\ }\href
  {https://doi.org/10.1126/science.aar6939} {\bibfield  {journal} {\bibinfo
  {journal} {Science}\ }\textbf {\bibinfo {volume} {360}},\ \bibinfo {pages}
  {764} (\bibinfo {year} {2018})}\BibitemShut {NoStop}%
\bibitem [{\citenamefont {Ren}\ \emph {et~al.}(2020)\citenamefont {Ren},
  \citenamefont {Matheny}, \citenamefont {MacCabe}, \citenamefont {Luo},
  \citenamefont {Pfeifer}, \citenamefont {Mirhosseini},\ and\ \citenamefont
  {Painter}}]{ren2020two}%
  \BibitemOpen
  \bibfield  {author} {\bibinfo {author} {\bibfnamefont {H.}~\bibnamefont
  {Ren}}, \bibinfo {author} {\bibfnamefont {M.~H.}\ \bibnamefont {Matheny}},
  \bibinfo {author} {\bibfnamefont {G.~S.}\ \bibnamefont {MacCabe}}, \bibinfo
  {author} {\bibfnamefont {J.}~\bibnamefont {Luo}}, \bibinfo {author}
  {\bibfnamefont {H.}~\bibnamefont {Pfeifer}}, \bibinfo {author} {\bibfnamefont
  {M.}~\bibnamefont {Mirhosseini}},\ and\ \bibinfo {author} {\bibfnamefont
  {O.}~\bibnamefont {Painter}},\ }\bibfield  {title} {\bibinfo {title}
  {Two-dimensional optomechanical crystal cavity with high quantum
  cooperativity},\ }\href {https://doi.org/10.1038/s41467-020-17182-9}
  {\bibfield  {journal} {\bibinfo  {journal} {Nat. Commun.}\ }\textbf {\bibinfo
  {volume} {11}},\ \bibinfo {pages} {3373} (\bibinfo {year}
  {2020})}\BibitemShut {NoStop}%
\bibitem [{\citenamefont {Postma}\ \emph {et~al.}(2005)\citenamefont {Postma},
  \citenamefont {Kozinsky}, \citenamefont {Husain},\ and\ \citenamefont
  {Roukes}}]{postma2005dynamic}%
  \BibitemOpen
  \bibfield  {author} {\bibinfo {author} {\bibfnamefont {H.~W.~C.}\
  \bibnamefont {Postma}}, \bibinfo {author} {\bibfnamefont {I.}~\bibnamefont
  {Kozinsky}}, \bibinfo {author} {\bibfnamefont {A.}~\bibnamefont {Husain}},\
  and\ \bibinfo {author} {\bibfnamefont {M.~L.}\ \bibnamefont {Roukes}},\
  }\bibfield  {title} {\bibinfo {title} {{Dynamic range of nanotube- and
  nanowire-based electromechanical systems}},\ }\href
  {https://doi.org/10.1063/1.1929098} {\bibfield  {journal} {\bibinfo
  {journal} {Appl. Phys. Lett.}\ }\textbf {\bibinfo {volume} {86}},\ \bibinfo
  {pages} {223105} (\bibinfo {year} {2005})}\BibitemShut {NoStop}%
\bibitem [{\citenamefont {Moser}\ \emph {et~al.}(2014)\citenamefont {Moser},
  \citenamefont {Eichler}, \citenamefont {G{\"u}ttinger}, \citenamefont
  {Dykman},\ and\ \citenamefont {Bachtold}}]{moser2014b}%
  \BibitemOpen
  \bibfield  {author} {\bibinfo {author} {\bibfnamefont {J.}~\bibnamefont
  {Moser}}, \bibinfo {author} {\bibfnamefont {A.}~\bibnamefont {Eichler}},
  \bibinfo {author} {\bibfnamefont {J.}~\bibnamefont {G{\"u}ttinger}}, \bibinfo
  {author} {\bibfnamefont {M.~I.}\ \bibnamefont {Dykman}},\ and\ \bibinfo
  {author} {\bibfnamefont {A.}~\bibnamefont {Bachtold}},\ }\bibfield  {title}
  {\bibinfo {title} {Nanotube mechanical resonators with quality factors of up
  to 5 million},\ }\href {https://doi.org/10.1038/nnano.2014.234} {\bibfield
  {journal} {\bibinfo  {journal} {Nat. Nanotech.}\ }\textbf {\bibinfo {volume}
  {9}},\ \bibinfo {pages} {1007} (\bibinfo {year} {2014})}\BibitemShut
  {NoStop}%
\bibitem [{\citenamefont {Chardin}\ \emph {et~al.}(2024)\citenamefont
  {Chardin}, \citenamefont {Pairis}, \citenamefont {Douillet}, \citenamefont
  {Hocevar}, \citenamefont {Claudon}, \citenamefont {Poizat}, \citenamefont
  {Bellon},\ and\ \citenamefont {Verlot}}]{chardin2024}%
  \BibitemOpen
  \bibfield  {author} {\bibinfo {author} {\bibfnamefont {C.}~\bibnamefont
  {Chardin}}, \bibinfo {author} {\bibfnamefont {S.}~\bibnamefont {Pairis}},
  \bibinfo {author} {\bibfnamefont {S.}~\bibnamefont {Douillet}}, \bibinfo
  {author} {\bibfnamefont {M.}~\bibnamefont {Hocevar}}, \bibinfo {author}
  {\bibfnamefont {J.}~\bibnamefont {Claudon}}, \bibinfo {author} {\bibfnamefont
  {J.-P.}\ \bibnamefont {Poizat}}, \bibinfo {author} {\bibfnamefont
  {L.}~\bibnamefont {Bellon}},\ and\ \bibinfo {author} {\bibfnamefont
  {P.}~\bibnamefont {Verlot}},\ }\bibfield  {title} {\bibinfo {title}
  {Hyperspectral electromechanical imaging at the nanoscale: Dynamical
  backaction, dissipation and quantum fluctuations},\ }\bibfield  {journal}
  {\bibinfo  {journal} {arXiv:2407.20740}\ }\href
  {https://doi.org/10.48550/arXiv.2407.20740} {10.48550/arXiv.2407.20740}
  (\bibinfo {year} {2024})\BibitemShut {NoStop}%
\bibitem [{\citenamefont {Thijssen}\ \emph {et~al.}(2013)\citenamefont
  {Thijssen}, \citenamefont {Verhagen}, \citenamefont {Kippenberg},\ and\
  \citenamefont {Polman}}]{thijssen2013plasmon}%
  \BibitemOpen
  \bibfield  {author} {\bibinfo {author} {\bibfnamefont {R.}~\bibnamefont
  {Thijssen}}, \bibinfo {author} {\bibfnamefont {E.}~\bibnamefont {Verhagen}},
  \bibinfo {author} {\bibfnamefont {T.~J.}\ \bibnamefont {Kippenberg}},\ and\
  \bibinfo {author} {\bibfnamefont {A.}~\bibnamefont {Polman}},\ }\bibfield
  {title} {\bibinfo {title} {Plasmon nanomechanical coupling for nanoscale
  transduction},\ }\href {https://doi.org/10.1021/nl4015028} {\bibfield
  {journal} {\bibinfo  {journal} {Nano Lett.}\ }\textbf {\bibinfo {volume}
  {13}},\ \bibinfo {pages} {3293} (\bibinfo {year} {2013})}\BibitemShut
  {NoStop}%
\bibitem [{\citenamefont {Montinaro}\ \emph {et~al.}(2014)\citenamefont
  {Montinaro}, \citenamefont {W{\"u}st}, \citenamefont {Munsch}, \citenamefont
  {Fontana}, \citenamefont {Russo-Averchi}, \citenamefont {Heiss},
  \citenamefont {Fontcuberta~i Morral}, \citenamefont {Warburton},\ and\
  \citenamefont {Poggio}}]{montinaro2014quantum}%
  \BibitemOpen
  \bibfield  {author} {\bibinfo {author} {\bibfnamefont {M.}~\bibnamefont
  {Montinaro}}, \bibinfo {author} {\bibfnamefont {G.}~\bibnamefont {W{\"u}st}},
  \bibinfo {author} {\bibfnamefont {M.}~\bibnamefont {Munsch}}, \bibinfo
  {author} {\bibfnamefont {Y.}~\bibnamefont {Fontana}}, \bibinfo {author}
  {\bibfnamefont {E.}~\bibnamefont {Russo-Averchi}}, \bibinfo {author}
  {\bibfnamefont {M.}~\bibnamefont {Heiss}}, \bibinfo {author} {\bibfnamefont
  {A.}~\bibnamefont {Fontcuberta~i Morral}}, \bibinfo {author} {\bibfnamefont
  {R.~J.}\ \bibnamefont {Warburton}},\ and\ \bibinfo {author} {\bibfnamefont
  {M.}~\bibnamefont {Poggio}},\ }\bibfield  {title} {\bibinfo {title} {Quantum
  dot opto-mechanics in a fully self-assembled nanowire},\ }\href
  {https://doi.org/10.1021/nl501413t} {\bibfield  {journal} {\bibinfo
  {journal} {Nano Lett.}\ }\textbf {\bibinfo {volume} {14}},\ \bibinfo {pages}
  {4454} (\bibinfo {year} {2014})}\BibitemShut {NoStop}%
\bibitem [{\citenamefont {Yeo}\ \emph {et~al.}(2014)\citenamefont {Yeo},
  \citenamefont {De~Assis}, \citenamefont {Gloppe}, \citenamefont
  {Dupont-Ferrier}, \citenamefont {Verlot}, \citenamefont {Malik},
  \citenamefont {Dupuy}, \citenamefont {Claudon}, \citenamefont {G{\'e}rard},
  \citenamefont {Auff{\`e}ves}, , \citenamefont {Nogues}, \citenamefont
  {Seidelin}, \citenamefont {Poizat}, \citenamefont {Arcizet},\ and\
  \citenamefont {Richard}}]{yeo2014strain}%
  \BibitemOpen
  \bibfield  {author} {\bibinfo {author} {\bibfnamefont {I.}~\bibnamefont
  {Yeo}}, \bibinfo {author} {\bibfnamefont {P.-L.}\ \bibnamefont {De~Assis}},
  \bibinfo {author} {\bibfnamefont {A.}~\bibnamefont {Gloppe}}, \bibinfo
  {author} {\bibfnamefont {E.}~\bibnamefont {Dupont-Ferrier}}, \bibinfo
  {author} {\bibfnamefont {P.}~\bibnamefont {Verlot}}, \bibinfo {author}
  {\bibfnamefont {N.~S.}\ \bibnamefont {Malik}}, \bibinfo {author}
  {\bibfnamefont {E.}~\bibnamefont {Dupuy}}, \bibinfo {author} {\bibfnamefont
  {J.}~\bibnamefont {Claudon}}, \bibinfo {author} {\bibfnamefont {J.-M.}\
  \bibnamefont {G{\'e}rard}}, \bibinfo {author} {\bibfnamefont
  {A.}~\bibnamefont {Auff{\`e}ves}}, , \bibinfo {author} {\bibfnamefont
  {G.}~\bibnamefont {Nogues}}, \bibinfo {author} {\bibfnamefont
  {S.}~\bibnamefont {Seidelin}}, \bibinfo {author} {\bibfnamefont {J.-P.}\
  \bibnamefont {Poizat}}, \bibinfo {author} {\bibfnamefont {O.}~\bibnamefont
  {Arcizet}},\ and\ \bibinfo {author} {\bibfnamefont {M.}~\bibnamefont
  {Richard}},\ }\bibfield  {title} {\bibinfo {title} {Strain-mediated coupling
  in a quantum dot--mechanical oscillator hybrid system},\ }\href
  {https://doi.org/10.1038/nnano.2013.274} {\bibfield  {journal} {\bibinfo
  {journal} {Nat. Nanotech.}\ }\textbf {\bibinfo {volume} {9}},\ \bibinfo
  {pages} {106} (\bibinfo {year} {2014})}\BibitemShut {NoStop}%
\bibitem [{\citenamefont {Rugar}\ \emph {et~al.}(2004)\citenamefont {Rugar},
  \citenamefont {Budakian}, \citenamefont {Mamin},\ and\ \citenamefont
  {Chui}}]{rugar2004single}%
  \BibitemOpen
  \bibfield  {author} {\bibinfo {author} {\bibfnamefont {D.}~\bibnamefont
  {Rugar}}, \bibinfo {author} {\bibfnamefont {R.}~\bibnamefont {Budakian}},
  \bibinfo {author} {\bibfnamefont {H.}~\bibnamefont {Mamin}},\ and\ \bibinfo
  {author} {\bibfnamefont {B.}~\bibnamefont {Chui}},\ }\bibfield  {title}
  {\bibinfo {title} {Single spin detection by magnetic resonance force
  microscopy},\ }\href {https://doi.org/10.1038/nature02658} {\bibfield
  {journal} {\bibinfo  {journal} {Nature}\ }\textbf {\bibinfo {volume} {430}},\
  \bibinfo {pages} {329} (\bibinfo {year} {2004})}\BibitemShut {NoStop}%
\bibitem [{\citenamefont {Hao}\ \emph {et~al.}(2003)\citenamefont {Hao},
  \citenamefont {Erbil},\ and\ \citenamefont {Ayazi}}]{Hao2003}%
  \BibitemOpen
  \bibfield  {author} {\bibinfo {author} {\bibfnamefont {Z.}~\bibnamefont
  {Hao}}, \bibinfo {author} {\bibfnamefont {A.}~\bibnamefont {Erbil}},\ and\
  \bibinfo {author} {\bibfnamefont {F.}~\bibnamefont {Ayazi}},\ }\bibfield
  {title} {\bibinfo {title} {An analytical model for support loss in
  micromachined beam resonators with in-plane flexural vibrations},\ }\href
  {https://doi.org/https://doi.org/10.1016/j.sna.2003.09.037} {\bibfield
  {journal} {\bibinfo  {journal} {Sens. Actuator A-Phys.}\ }\textbf {\bibinfo
  {volume} {109}},\ \bibinfo {pages} {156} (\bibinfo {year}
  {2003})}\BibitemShut {NoStop}%
\bibitem [{\citenamefont {Rieger}\ \emph {et~al.}(2014)\citenamefont {Rieger},
  \citenamefont {Isacsson}, \citenamefont {Seitner}, \citenamefont {Kotthaus},\
  and\ \citenamefont {Weig}}]{rieger2014energy}%
  \BibitemOpen
  \bibfield  {author} {\bibinfo {author} {\bibfnamefont {J.}~\bibnamefont
  {Rieger}}, \bibinfo {author} {\bibfnamefont {A.}~\bibnamefont {Isacsson}},
  \bibinfo {author} {\bibfnamefont {M.~J.}\ \bibnamefont {Seitner}}, \bibinfo
  {author} {\bibfnamefont {J.~P.}\ \bibnamefont {Kotthaus}},\ and\ \bibinfo
  {author} {\bibfnamefont {E.~M.}\ \bibnamefont {Weig}},\ }\bibfield  {title}
  {\bibinfo {title} {Energy losses of nanomechanical resonators induced by
  atomic force microscopy-controlled mechanical impedance mismatching},\ }\href
  {https://doi.org/10.1038/ncomms4345} {\bibfield  {journal} {\bibinfo
  {journal} {Nat. Commun.}\ }\textbf {\bibinfo {volume} {5}},\ \bibinfo {pages}
  {3345} (\bibinfo {year} {2014})}\BibitemShut {NoStop}%
\bibitem [{\citenamefont {Morell}\ \emph {et~al.}(2019)\citenamefont {Morell},
  \citenamefont {Tepsic}, \citenamefont {Reserbat-Plantey}, \citenamefont
  {Cepellotti}, \citenamefont {Manca}, \citenamefont {Epstein}, \citenamefont
  {Isacsson}, \citenamefont {Marie}, \citenamefont {Mauri},\ and\ \citenamefont
  {Bachtold}}]{Morell-2019}%
  \BibitemOpen
  \bibfield  {author} {\bibinfo {author} {\bibfnamefont {N.}~\bibnamefont
  {Morell}}, \bibinfo {author} {\bibfnamefont {S.}~\bibnamefont {Tepsic}},
  \bibinfo {author} {\bibfnamefont {A.}~\bibnamefont {Reserbat-Plantey}},
  \bibinfo {author} {\bibfnamefont {A.}~\bibnamefont {Cepellotti}}, \bibinfo
  {author} {\bibfnamefont {M.}~\bibnamefont {Manca}}, \bibinfo {author}
  {\bibfnamefont {I.}~\bibnamefont {Epstein}}, \bibinfo {author} {\bibfnamefont
  {A.}~\bibnamefont {Isacsson}}, \bibinfo {author} {\bibfnamefont
  {X.}~\bibnamefont {Marie}}, \bibinfo {author} {\bibfnamefont
  {F.}~\bibnamefont {Mauri}},\ and\ \bibinfo {author} {\bibfnamefont
  {A.}~\bibnamefont {Bachtold}},\ }\bibfield  {title} {\bibinfo {title}
  {Optomechanical measurement of thermal transport in two-dimensional mose2
  lattices},\ }\href {https://doi.org/10.1021/acs.nanolett.9b00560} {\bibfield
  {journal} {\bibinfo  {journal} {Nano Lett.}\ }\textbf {\bibinfo {volume}
  {19}},\ \bibinfo {pages} {3143} (\bibinfo {year} {2019})}\BibitemShut
  {NoStop}%
\bibitem [{\citenamefont {Iadanza}\ \emph {et~al.}(2020)\citenamefont
  {Iadanza}, \citenamefont {Clementi}, \citenamefont {Hu}, \citenamefont
  {Schulz}, \citenamefont {Gerace}, \citenamefont {Galli},\ and\ \citenamefont
  {O'Faolain}}]{Iadanza-2020}%
  \BibitemOpen
  \bibfield  {author} {\bibinfo {author} {\bibfnamefont {S.}~\bibnamefont
  {Iadanza}}, \bibinfo {author} {\bibfnamefont {M.}~\bibnamefont {Clementi}},
  \bibinfo {author} {\bibfnamefont {C.}~\bibnamefont {Hu}}, \bibinfo {author}
  {\bibfnamefont {S.~A.}\ \bibnamefont {Schulz}}, \bibinfo {author}
  {\bibfnamefont {D.}~\bibnamefont {Gerace}}, \bibinfo {author} {\bibfnamefont
  {M.}~\bibnamefont {Galli}},\ and\ \bibinfo {author} {\bibfnamefont
  {L.}~\bibnamefont {O'Faolain}},\ }\bibfield  {title} {\bibinfo {title} {Model
  of thermo-optic nonlinear dynamics of photonic crystal cavities},\ }\href
  {https://doi.org/10.1103/PhysRevB.102.245404} {\bibfield  {journal} {\bibinfo
   {journal} {Phys. Rev. B}\ }\textbf {\bibinfo {volume} {102}},\ \bibinfo
  {pages} {245404} (\bibinfo {year} {2020})}\BibitemShut {NoStop}%
\bibitem [{\citenamefont {Unterreithmeier}\ \emph {et~al.}(2009)\citenamefont
  {Unterreithmeier}, \citenamefont {Weig},\ and\ \citenamefont
  {Kotthaus}}]{unterreithmeier2009universal}%
  \BibitemOpen
  \bibfield  {author} {\bibinfo {author} {\bibfnamefont {Q.~P.}\ \bibnamefont
  {Unterreithmeier}}, \bibinfo {author} {\bibfnamefont {E.~M.}\ \bibnamefont
  {Weig}},\ and\ \bibinfo {author} {\bibfnamefont {J.~P.}\ \bibnamefont
  {Kotthaus}},\ }\bibfield  {title} {\bibinfo {title} {Universal transduction
  scheme for nanomechanical systems based on dielectric forces},\ }\href
  {https://doi.org/10.1038/nature07932} {\bibfield  {journal} {\bibinfo
  {journal} {Nature}\ }\textbf {\bibinfo {volume} {458}},\ \bibinfo {pages}
  {1001} (\bibinfo {year} {2009})}\BibitemShut {NoStop}%
\bibitem [{\citenamefont {Khivrich}\ \emph {et~al.}(2019)\citenamefont
  {Khivrich}, \citenamefont {Clerk},\ and\ \citenamefont
  {Ilani}}]{khivrich2019nanomechanical}%
  \BibitemOpen
  \bibfield  {author} {\bibinfo {author} {\bibfnamefont {I.}~\bibnamefont
  {Khivrich}}, \bibinfo {author} {\bibfnamefont {A.~A.}\ \bibnamefont
  {Clerk}},\ and\ \bibinfo {author} {\bibfnamefont {S.}~\bibnamefont {Ilani}},\
  }\bibfield  {title} {\bibinfo {title} {Nanomechanical pump--probe
  measurements of insulating electronic states in a carbon nanotube},\ }\href
  {https://doi.org/10.1038/s41565-018-0341-6} {\bibfield  {journal} {\bibinfo
  {journal} {Nat. Nanotech.}\ }\textbf {\bibinfo {volume} {14}},\ \bibinfo
  {pages} {161} (\bibinfo {year} {2019})}\BibitemShut {NoStop}%
\bibitem [{\citenamefont {Gouriou}(2023)}]{Gouriou2023PhDThesis}%
  \BibitemOpen
  \bibfield  {author} {\bibinfo {author} {\bibfnamefont {C.}~\bibnamefont
  {Gouriou}},\ }\emph {\bibinfo {title} {{Nano-optomechanics of suspended SiC
  nanowires down to cryogenic temperatures : exploration of optical and
  photothermal responses}}},\ \href {https://theses.hal.science/tel-04444724}
  {Ph.D. thesis},\ \bibinfo  {school} {{Universit{\'e} Grenoble Alpes}}
  (\bibinfo {year} {2023})\BibitemShut {NoStop}%
\end{thebibliography}%

\end{document}